\documentclass[aps,prb,reprint,superscriptaddress]{revtex4-2}

\usepackage[english]{babel}
\usepackage[utf8x]{inputenc}
\usepackage[T1]{fontenc}
\usepackage{siunitx}
\usepackage{xcolor}
\usepackage[version=4]{mhchem}

\usepackage{longtable}
\usepackage{booktabs}
\usepackage{bibentry}
\usepackage{amsmath}
\usepackage{graphicx}
\usepackage[nolist]{acronym}
\usepackage{xspace}
\usepackage{bm}
\usepackage{physics}
\usepackage{xr}
\newcommand{\gstate}{${}^4$I$_{15/2}$\xspace}
\newcommand{\estate}{${}^4$I$_{13/2}$\xspace}
\newcommand{\ert}{Er$^{3+}$\xspace}
\newcommand{\tplus}{$^{3+}$\xspace}
\newcommand{\yso}{Y$_2$SiO$_5$\xspace}
\newcommand{\ertrans}{${}^4$I$_{15/2}$ $\rightarrow$ ${}^4$I$_{13/2}$\xspace}
\newcommand{\iso}[1]{$^{#1}$}

\begin{document}

\title{Effect of a hybrid transition moment on Stark-modulated photon echoes in Er\tplus:\yso}

\author{Rose L. \surname{Ahlefeldt}}
\email{rose.ahlefeldt@anu.edu.au}
\affiliation{Centre of Excellence for Quantum Computation and Communication Technology, Research School of Physics, Australian National University, Canberra, ACT 0200, Australia}
\author{Alexey \surname{Lyasota}}
\affiliation{Centre of Excellence for Quantum Computation and Communication Technology, School of Physics, University of New South Wales, Sydney, NSW 2052, Australia}
\author{Jodie \surname{Smith}}
\affiliation{Centre of Excellence for Quantum Computation and Communication Technology, Research School of Physics, Australian National University, Canberra, ACT 0200, Australia}
\author{Jinliang \surname{Ren}}
\affiliation{Centre of Excellence for Quantum Computation and Communication Technology, Research School of Physics, Australian National University, Canberra, ACT 0200, Australia}
\author{Matthew J. \surname{Sellars}}
\affiliation{Centre of Excellence for Quantum Computation and Communication Technology, Research School of Physics, Australian National University, Canberra, ACT 0200, Australia}



\begin{abstract}
The 1538~nm \ertrans transition of \ert has an unusual hybrid electric-magnetic dipole character, and signatures of the hybrid moment can be expected in coherent transient measurements. Here, we investigate the effect of the hybrid moment in both sites of \ert:\yso on Stark-modulated photon echo measurements, showing that it results in a reduction of  visibility of the modulated signal as well as phase and polarization shifts. We interpret these effects using a simple optical Bloch equation model, showing that site 1 has a strongly mixed moment and site 2 is predominantly magnetic dipole. We discuss the implications of the hybrid moment for quantum information applications of quantum memories. We also use the echo measurements to extract the Stark shift of the optical transition along three orthogonal directions, finding values between 10.50 and \SI{11.93}{\kilo\hertz\per(\volt\per\cm)} for site 1 and 1.61 and \SI{15.35}{\kilo\hertz\per(\volt\per\cm)} for site 2. We observe a modification of the Zeeman shift by the electric field in both sites and discuss how this may be used to  electrically control \ert spin qubits.
\end{abstract}

\maketitle
\section{Introduction}

Crystals containing \ert ions have emerged as a particularly attractive candidate for quantum information applications that require flying qubits. In addition to the long optical and hyperfine coherence times commonly seen in the rare earth ions, \ert has an optical transition from its \gstate ground state near \SI{1540}{nm}, in the middle of the telecommunications C-band, allowing control and communication between \ert-based devices at a wavelength compatible with existing optical fibre technology. As a result, \ert materials are currently studied for quantum computing, quantum memories, optical quantum interconnects and quantum repeaters \cite{rochman23,chen20,gritsch22,berkman23,stuart21}.

The 1540~nm \gstate -- \estate transition of \ert has a less well-known property, almost unique amongst rare earth transitions studied for quantum information: in addition to the normal forced electric dipole transition moment arising in sites with a non-centrosymmetric crystal field, it has a magnetic  dipole transition moment comparable to the electric dipole in most \ert crystals \cite{weber67, weber68, weber73, li14c, gerasimov16, Xie21}. 

The response of a crystal with a hybrid moment under coherent excitation can be fundamentally different from that with only an electric dipole moment, due to the different properties of electric and magnetic dipole moments under inversion symmetry. Nearly every \ert crystal studied for quantum information, including the \ert:\yso crystal studied here, has a crystal structure with inversion symmetry, but with the \ert ion located at a site without inversion symmetry. In these materials, then, each crystallographic site occupied by \ert is composed of two inversion-related sub-sites. The dipole moments of these sites are also related by inversion, but electric and magnetic dipoles are, respectively, odd and even under inversion. The consequence is that, when both moments are comparable, the inversion-related sub-sites can be expected to have \textit{different} interactions with light: different Rabi frequencies, different polarization dependence, and so on. 


These differences can be seen in experiments with applied electric fields, and here we report measurements using a Stark-modulated photon echo method \cite{wang92, meixner92,macfarlane94a, graf97}. In this technique, an electric field is applied during part of the wait time in a photon echo sequence. The odd symmetry of the permanent electric dipole moment means the inversion-related sites accumulate opposite phase while the field is on, resulting in interference between the two sites dependent on the accumulated phase, and therefore a modulation in the echo amplitude as a function of the field. 

We originally took Stark-modulated photon echo measurements to determine the Stark shift of \ert:\yso, which has only been partially determined \cite{craiciu21,minar09}. These measurements are necessary because electric fields are commonly used to control atomic coherence in a variety of quantum memory and quantum information applications\cite{hetet08, damon11, mcauslan11, craiciu21}. However, we saw that the modulated signal was very different from that expected for an electric dipole transition. Here we interpret the modulated signals taking into account the hybrid moment using a simple optical Bloch equation model. We use the Stark modulated photon echo to investigate the hybrid moment of both site 1 and site 2 in \ert:\yso, reporting at the same time the Stark shifts of both sites. We show that both sites display a stark shift of the Zeeman splitting (a $g$-shift) in addition to the optical Stark shift. Finally, we discuss some of the consequences the hybrid moment and the Stark shifts seen have for quantum information applications of \ert:\yso and other similar crystals.



\section{Optical Bloch equation description}\label{OBE}
Photon echoes in dilute media are typically well explained by a Maxwell-Bloch equation model describing the atomic polarization coupled to a classical light field. Numerical implementations of these models can be used to accurately model experimental data. However, \ert:\yso has three complications not considered in existing numerical models: 1) it is biaxial birefringent, 2) the transition is both electric- and magnetic-dipole allowed, and 3) these moments are not constrained to lie along the same directions as each other or along the principal axes of the anisotropic refractive index tensor. Since the refractive index tensor has only been characterized at visible frequencies \cite{beach90a}, and the electric and magnetic dipole moments are only partially known \cite{petit20, sun05}, the Maxwell Bloch model has a large number of free parameters in \ert:\yso. 

 Here, we aim only to describe the general behavior of the Stark-modulated photon echo signal in the case of a hybrid transition moment. We use a simpler optical Bloch equation model, which ignores the effect the atoms have on the propagating light field. We will also ignore the birefringence, as the resulting analytical equations provide clearer insight into the effects of the hybrid moment.  This model is presented in many textbooks, and here we briefly describe the modifications to this model required to account for the mixed dipole moment, following the nomenclature of Levenson and Kano \cite{levenson89}.

Consider a two-level atom described by states $\ket{a}$ and $\ket{b}$ coupled to a semi-classical light field. The system Hamiltonian is:
\begin{equation}
H = H_0+H_1+H_R \label{systemH}
\end{equation}
where $H_0$ is the bare atom Hamiltonian, $H_1$ describes the interaction with light, and $H_R$ describes relaxation. $H_1$ can be written in the density matrix formalism
\begin{align}
    H_1 = \begin{bmatrix}0&H_{ba}^*\\H_{ba}&0\end{bmatrix}\,,
\end{align}
with
\begin{align}
        H_{ba} &= -\bra{b}\bm{d}\ket{a}\cdot\bm{E}(t) -\bra{b}\bm{m}\ket{a}\cdot\bm{B}(t)\,,\\
         &= -\bm{d}_{ba}\cdot\bm{E}(t) -\bm{m}_{ba}\cdot\bm{B}(t)\,,
\end{align}
where $\bm{d}_{ba}$ is the electric dipole matrix element between states $b$ and $a$, $\bm{m}_{ba}$ is the corresponding magnetic dipole moment matrix element, and the electromagnetic field at frequency $\omega_0$ is defined by
\begin{align}
    \bm{E}(t) &= \frac{1}{2}E_0(t)\bm{\hat{\varepsilon}}e^{-i\omega_0 t}+\mathrm{c.c.}\label{efield} \,,\\
\bm{B}(t) &= \frac{n}{2c}E_0(t)(\bm{\hat{k}\times\hat{\varepsilon}})e^{-i\omega_0 t} +\mathrm{c.c.}\label{bfield}\,.
\end{align}
 Here, $\hat{\bm{\varepsilon}}$ is a unit vector describing the polarization of the light, $\bm{\hat{k}}$ is the unit wavevector, and we have ignored the spatial component of the light. We have written the magnetic field in terms of the electric field amplitude $E_0$ via the refractive index $n$ and the speed of light $c$. Under the rotating wave approximation (RWA), the interaction can be written 
\begin{align}
    H_1 = \hbar\begin{bmatrix}0&\chi^* e^{+i\omega_0 t}\\ \chi e^{-i\omega_0 t}&0\end{bmatrix}\,
    \label{H1}
\end{align}
with the complex Rabi frequency $\chi$ dependent on a total $k$-dependent dipole moment $\bm{\mu}_{ba}(\bm{\hat{k}})$ according to:
\begin{align}
    \chi &= -\frac{E_0(t)}{\hbar}\left[\bm{d}_{ba}\cdot \bm{\hat{\varepsilon}}+\frac{1}{v}\bm{m}_{ba}\cdot (\bm{\hat{k}\times\hat{\varepsilon}})\right]\\
    & = -\frac{E_0(t)}{\hbar}\left[\bm{d}_{ba} +\frac{n}{c} (\bm{m}_{ba}\times\bm{\hat{k}})\right]\cdot \bm{\hat{\varepsilon}}\\
    & = E_0(t)\bm{\mu}_{ba}(\bm{\hat{k}})\cdot \bm{\hat{\varepsilon}}
\end{align}

Eq. \eqref{H1} is identical to the electric-dipole-only form of the interaction Hamiltonian in the standard formulation of the optical Bloch equations, so the time evolution of the density matrix under the system Hamiltonian Eq. \eqref{systemH} is that provided in many textbooks:
\begin{align}
\dot{\rho}_{aa} &= -\frac{i}{2}\left[\rho_{ba}\chi^*(z,t)e^{+i\omega_0t}-\rho_{ab}\chi(z,t)e^{-i\omega_0t}\right] +\frac{1}{T_1}(1-\rho_{aa})\nonumber \\
\dot{\rho}_{bb} &= +\frac{i}{2}\left[\rho_{ba}\chi^*(z,t)e^{+i\omega_0t}-\rho_{ab}\chi(z,t)e^{-i\omega_0t}\right] -\frac{1}{T_1}\rho_{bb}\nonumber \\
\dot{\rho}_{ab} &= -\frac{i}{2}(\rho_{bb}-\rho_{aa})\chi^*(z,t)e^{+i\omega_0 t} +i\Omega\rho_{ab} -\frac{1}{T_2}\rho_{ab}\nonumber \\
\dot{\rho}_{ba} &= \frac{i}{2}(\rho_{bb}-\rho_{aa})\chi(z,t)e^{-i\omega_0 t} -i\Omega\rho_{ba} -\frac{1}{T_2}\rho_{ba} 
\label{masterd}
\end{align}
where we have used the phenomenological decay constants $T_1$ and $T_2$ to describe relaxation processes. The density matrix can be related to the field radiated by the atoms by a polarization $\bm{D} = N\mathrm{Tr}(\bm{d}\rho)$ and a magnetization  $\bm{M} = N\mathrm{Tr}(\bm{m}\rho)$ with $N$ the number density of the material. In the absence of birefringence, we can define a total polarization that combines the electric polarization and the magnetization,
\begin{equation}
    \bm{P} = N\mathrm{Tr}(\bm{\mu}(\bm{\hat{k}})\rho)
    \label{polarization}
\end{equation}
that replaces $\bm{D}$ in the equations for the electric-dipole-only case. 

Interesting effects arise when solving Eqs. \eqref{masterd} and \eqref{polarization} for  non-centrosymmetric sites in centrosymmetric crystals.  In this case, the observed field radiated by the atom is the coherent sum of contributions from each inversion-related sub-site. Since electric $\bm{d}_{ba}$ and magnetic $\bm{m}_{ba}$ dipoles are odd and even under inversion, respectively, when both moments are non zero the total moment $\bm{\mu}$ is, in general, different for the two sub-sites. In the most extreme case, if $\frac{n}{c}\bm{\hat{k}}\cross\bm{m}_{ba} = \bm{d}_{ba}$, one sub-site will have a moment $\bm{\mu}_1(\bm{\hat{k}}) = 2\bm{d}_{ba}$ and the other no moment, $\bm{\mu}_2(\bm{\hat{k}}) = 0$, as the two contributions cancel. 

There are several obvious consequences of these unequal moments. First, the two sub-sites have different Rabi frequencies and will be driven to different positions on the Bloch sphere by the optical field, negatively affecting the ensemble state fidelity unless moment differences are small and the pulse sequence applied is robust to variations in Rabi frequency. Second, the two sub-sites will contribute unevenly to a radiated signal since the radiated field is proportional to $\bm{\mu}_{ba}$. This is important in experiments that rely on the interference of these subgroups, such as the Stark-modulated photon echoes here. Finally, the dependence of $\bm{\mu}_{ba}$ on $\bm{\hat{k}}$ means that the interaction of light with a single sub-site will be different in the forward and backward directions. To observe this non-reciprocal absorption, the  inversion symmetry must be broken by applying an electric field to spectrally resolve the two sub-sites so that only one can be excited. 

Here, we focus on the impact of a mixed dipole moment in Stark-modulated photon echo experiments \cite{wang92, meixner92}. This method, used to sensitively measure Stark shifts, involves applying a normal two-pulse photon echo to the transition of interest, but with an electric field pulse applied during part of the waiting time. For instance, the field may be applied only during the interval between the first and second pulses. In a material with inversion-related sub-sites, this field causes an opposite frequency shift on the two sub-sites, due, again, to the odd symmetry of electric fields under inversion symmetry, causing them to accrue equal and opposite phase. Since the field is not present in the interval between the second pulse and the echo, this phase accumulation is not reversed, and the interference of the two sub-sites reduces the final echo amplitude by an amount dependent on the accumulated phase; for a phase shift of $\frac{\pi}{2}$ the echo can be completely annihilated. Changing the field amplitude or pulse length, thus, results in a modulation of the echo amplitude that can be used to extract the transition's Stark shift.

Previous experimental Stark-modulated photon echo spectroscopy, reviewed by Macfarlane in 2007 \cite{macfarlane07}, has been performed on transitions that are electric-dipole allowed only. In this case, the sub-site moments are equal and opposite, and the echo amplitude undergoes a modulation with 100\% visibility  and a frequency given by twice the Stark shift $\Delta_s$ (strictly, this is true under certain, albeit commonly met, conditions only, see Supplemental Materials Fig. S1). 

In contrast, a mixed dipole moment leads to much more varied behavior of the modulations. To illustrate this fact, we numerically model the appropriate optical Bloch equations for a Stark-modulated photon echo sequence driving a mixed dipole moment in a material with inversion-related sub-sites. We considered square driving pulses short compared to the relaxation time, in which case both the pulses and the free evolution can be described by evolution matrices acting in sequence on an initial Bloch vector. These matrices can be derived from the optical Bloch equations presented in Eq. \eqref{masterd} and Ref. \onlinecite{levenson89}, taking care to use the general form for the  driven evolution that does not make the common assumption that the Rabi frequency is real.

 We modeled driving a flat inhomogeneous profile consisting of two inversion-related sub-sites 1 and 2 with, for sub-site 1, a 1~MHz Rabi frequency  and perfect, square $\pi$ and $\pi/2$ pulses ($t_\pi=\pi$~us) separated by $t_w = 13$~us. This is not a perfect $\pi$ pulse for sub-site 2 since it has, in general, a different Rabi frequency to sub-site 1. We drove the atoms with an electric field for $0\leq t_s \leq t_w$, causing a 0.5~MHz Stark shift on for $0\leq t_s \leq t_w$. To avoid any spurious modulations arising from the pulses driving the edges of our inhomogeneous distribution, we simulated for 5000 detunings between $-40$ and $+40$~MHz. We modeled a detection system analogous to that used in the experiment (Sec. \ref{exp}), where the output echo is split by a polarising beamsplitter into vertical and horizontal components whose intensity is detected on two different detectors. The light polarization was chosen along $\bm{\hat{x}}$ with $\bm{\hat{k}}$ along $\bm{\hat{z}}$. 

\begin{figure}
	\centering
    \includegraphics[width=1.0\columnwidth]{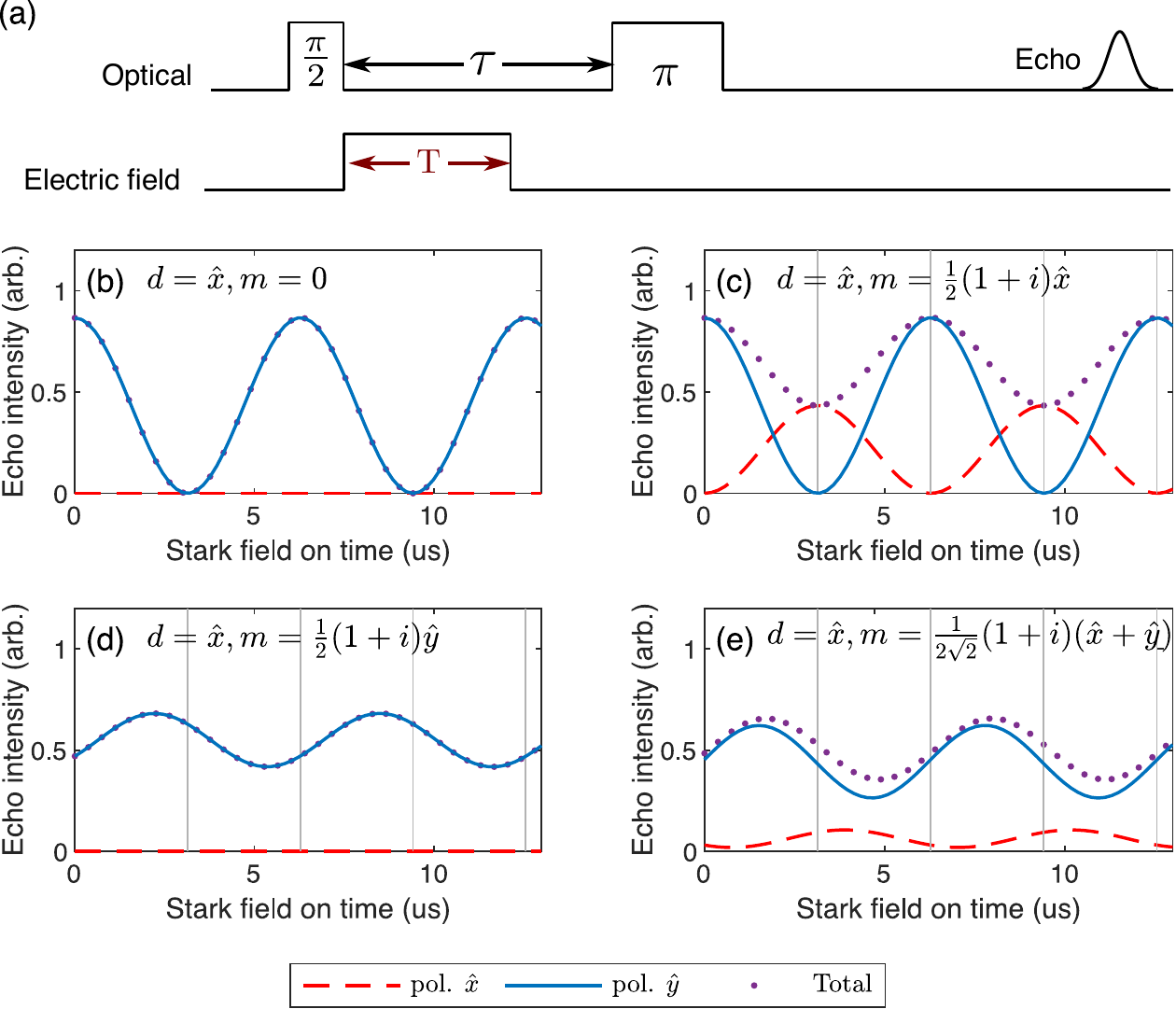}
    \caption{\label{modexamples} a) Stark modulated photon echo sequence. b)-e) Simulated peak echo intensity for the sequence in a) as a function of Stark field on time $T$, for an electric dipole moment $\bm{\hat{d}} = 1\bm{\hat{x}}$ and four different choices of magnetic dipole moment $\bm{\hat{m}}$, annotated on the plot: b)$\bm{\hat{m}} = 0$; c)  $\bm{\hat{m}}$ parallel to $\bm{\hat{d}}$; d) $\bm{\hat{m}}$ perpendicular to $\bm{\hat{d}}$; e) general $\bm{\hat{m}}$. In all figures, the blue line gives the signal seen in horizontal polarization, the red dashed line that in vertical polarization, and the purple dots the total signal. Grey lines indicate the positions of the peaks and troughs of the unshifted signal (b). }
\end{figure}
Representative $\hat{\bm{x}}$, $\hat{\bm{y}}$, and total intensity signals are shown in Fig. \ref{modexamples}, which plots $|\bm{P}|^2$, proportional to the peak echo intensity. In all cases, the electric dipole moment is $\bm{d} = 1\bm{\hat{x}}$~MHz/(V/cm), and the magnetic dipole moment was always smaller than the electric dipole. Four magnetic dipole cases are shown: (b) No $\bm{\hat{m}}$, (c) $\bm{\hat{m}}$ parallel to $\bm{\hat{d}}$, (d) $\bm{\hat{m}}$ perpendicular to $\bm{\hat{d}}$, and (e)  a general $\bm{\hat{m}}$. In (b), no magnetic dipole, the normal Stark modulated signal is seen: 100\% visibility modulations in $\hat{\bm{x}}$ polarization only with a frequency given by twice the Stark shift $\Delta_s$. In (c), parallel moments, the Rabi frequency is not modified from the previous case (since the magnetic dipole contribution to the moment is perpendicular to the light field polarization), but the radiated signal (directly dependent on the moment) is. The primary effect, then, is a polarization rotation of the output signal dependent on the field and the relative $\bm{\hat{d}}$ and $\bm{\hat{m}}$ contributions, giving full visibility modulations in the polarised detectors but not in the total signal. In (d), perpendicular moments, we see a shift in the phase combined with a reduction in the visibility of the output signal, while in (e), the general case, reduced visibility and phase shifts are seen on all three signals, combined with polarization rotation. These three characteristics -- phase shifts of the Stark modulation, reduction in visibility, and rotation of the output polarization even when $\bm{\hat{d}}||\bm{\hat{\varepsilon}}$ -- can be considered signatures of having a mixed moment.

The optical Bloch picture above is only strictly applicable to low optical-depth, isotropic crystals because it ignores absorption and birefringence. Birefringence, in particular, is common in materials that contain inversion-related sub-sites, including in \yso. However, the general features described above can be expected in birefringent media, and, indeed, none of the effects we have attributed to the mixed dipole moment are present in experimental studies of birefringent media with electric-dipole-only transitions.

If birefringence is included, a full Maxwell-Bloch model is needed because the optical Bloch model does not describe the evolving polarization of the propagating light field. This is true even when the light propagates along principal axes of the refractive index tensor because the atoms radiate in the direction of their dipole moments, which are not, in general, aligned with the refractive index tensor. Further, one can no longer write a combined dipole moment $\bm{\mu}$, so the Maxwell-Bloch equations have to be re-derived with birefringence and the hybrid moment considered. We will present such a re-derivation in later work.  

\section{Experimental Method} \label{exp}
We compare the general behavior seen in the previous section with the spectra obtained in \ert:\yso. \yso is a monoclinic crystal (space group I2/a) and \ert substitutes for Y\tplus in two C$_1$ symmetry crystallographic sites, known as site 1 and site 2 \cite{sun08}. The \ertrans Z$_1$ $\rightarrow$ Y$_1$ transition between the lowest crystal field doublets for these sites occurs at \SI{195.1}{THz} and \SI{194.8}{THz} for site 1 and 2 respectively. 

All measurements were performed on a 0.005\% \iso{167}Er\tplus:\yso crystal (from Scientific Materials) cut along the principal axes of the refractive index tensor, D$_1$, D$_2$, and C$_2$, with dimensions of $3.17\times4.15\times$\SI{5.15}{mm} ($\textrm{D}_1\times\textrm{D}_2\times\textrm{C}_2$). 

To accurately determine the (rather small) Stark shifts seen here using Stark-modulated photon echoes, we required a coherence time of the order of 100~us. The optical coherence time can be increased over the $\mathcal{O}(4)$~us zero-field value by applying a moderate magnetic field to reduce coupling to the nuclear spin bath \cite{petit20, fraval04}. We obtained coherence times of $>100$~us on both site 1 and 2 by cooling the crystal to 2~K in pumped liquid helium with a magnetic field of the order of 300~mT applied along the D$_1$ axis using two 25~mm diameter magnets on either side  of the sample (four magnets in total) and separated by approximately 10~mm. 

Here, we present measurements of the Stark shift for fields along all three optical extinction axes that sample all combinations of propagation and polarisation directions while maintaining the magnetic field along D$_1$. Two different sets of magnets were used to access these directions: for measurements with laser propagation  along D$_2$ and C$_2$, solid 3mm thick magnets were used, and for D$_1$ propagation, 4.5 mm magnets with a 5 mm hole in the center were used to allow passage of the laser beam. As a result, slightly different magnetic fields were present in each of the three directions, all of order 300~mT.

For each measurement direction, one pair of faces of the crystal were painted with silver paint and clamped between a pair of 10~mm diameter, 1~mm thick copper plates connected to an external voltage source supplying voltages up to \SI{10}{\V}. To get a homogeneous electric field in the sample, we ensured that the silver paint covered the entire surface of the perpendicular faces but did not extend at all onto the other faces and that the clamp applied sufficient pressure to ensure good electrical contact. Even with this effort, some of the spectra showed signs of inhomogeneity in the electric field (discussed later), due to tilted or misaligned plates. For all directions, we checked if the Stark shift was linear by measuring it at 2.5~V increments up to 10~V.  All Stark shifts were linear to the precision of our measurements so the values and spectra presented below are derived from the 10~V measurements only. 

The atoms were driven with approximately 7~mW of focused laser light supplied by a Pure Photonics PPCL560 tunable diode laser and gated by a single acousto-optic modulator. The laser polarization was purified by a polarising beamsplitter and the transmitted beam was focused with a \SI{10}{cm} lens onto the sample, with a half waveplate used to control the polarization. A second polarising beamsplitter was used to split the output echo into horizontal and vertical polarizations which were detected on two InGaAs photodiodes. 

\begin{figure}
	\centering
    \includegraphics[width=1.0\columnwidth]{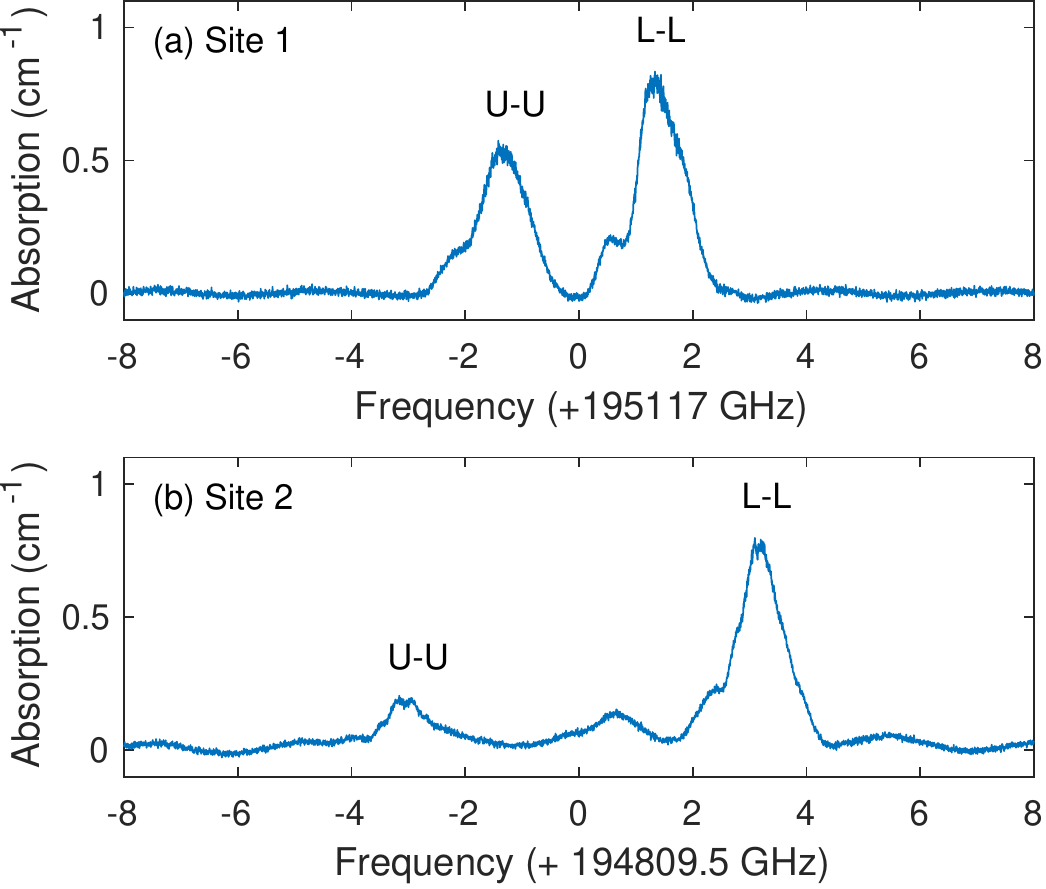}
    \caption{\label{spectra} Absorption spectra of \ert\yso for light propagating along D$_2$ at 2~K and $\approx 300$~mT for (a) site 1, with light polarized along C$_2$ and (b) site 2 for light polarized along D$_1$. Two peaks corresponding to lower-to-lower (L-L) and upper-to-upper (U-U) Zeeman transitions are shown. Additional hyperfine structure is seen on the L-L branch. The U-U peak is weaker because the Boltzmann population of this level is lower. The frequency scale is the uncalibrated internal reference of the laser. These spectra were obtained from the raw transmission data by dividing out large interference fringes arising from reflections off the  acousto-optic modulator (which was anti-reflection coated for visible frequencies, not 1550~nm).}
\end{figure}
The $\mathcal{O}(\SI{300}{\milli\tesla})$ field split the observed spectrum of both Er\tplus sites into two peaks separated by several gigahertz and  arising from the two like-to-like transitions between the \gstate and \estate doublets (Fig. \ref{spectra}): the lower-to-lower and upper-to-upper, henceforth referred to as the lower and upper Zeeman transitions. We measured Stark shifts on both peaks. For each measurement direction and site, we first recorded the photon echo amplitude as a function of delay time with no electric field  (Fig. S2 in the Supplemental materials) to map the superhyperfine modulations in the echo amplitude still present due to the residual coupling to the nuclear spin bath \cite{petit20}. With this information, we identified a suitable delay, $\tau$, for Stark modulation measurements where the superhyperfine interference was low. We used the Stark modulation pulse sequence shown in Fig. \ref{modexamples} (a) with a fixed voltage of 10 V applied to the capacitor and an on-time that varied between 0.1 and $\tau-1.5$~\SI{}{\micro\second}, which ensured that the electric field pulse did not overlap the light pulses. We used a $\pi$-pulse time of between 0.5 and \SI {0.7}{\micro\second} chosen to optimize the echo amplitude for the polarization that gave the largest signal. To reduce the complications due to birefringence, we only considered input polarizations along the optical extinction axes (horizontal and vertical in our setup).

For each electric field pulse length, we recorded eight shots with a wait time of 240~ms between shots to minimize fluctuations caused by holeburning into the hyperfine structure. To extract the echo area, we averaged these shots and then integrated the area under the echo after subtracting a background determined from an average of the signal in an area of the trace where the echo was absent. 

\section{Experimental results}

\begin{figure*}
	\centering
    \includegraphics[width=2.0\columnwidth]{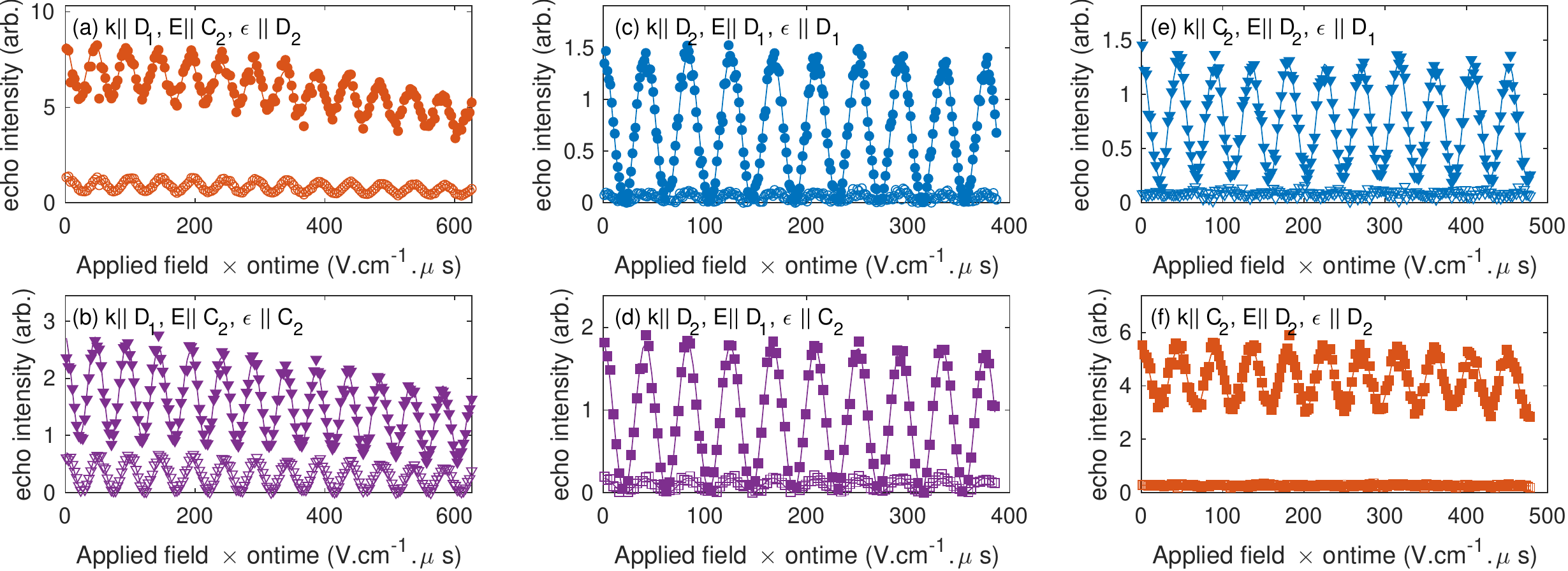}
    \caption{\label{S1mod} Stark echo modulations for site 1 \ert:\yso for three different combinations of propagation and applied electric field  directions, for two different input polarizations (linear along optical extinction axes) and two different detected polarizations (linear along optical extinction axes). (a) and (b), propagation along D$_1$, electric field along C$_2$. (c) and (d) propagation along D$_2$, electric field along D$_1$. (e) and (f), propagation along C$_2$, electric field along D$_2$. In all sub-figures, the orientation of the electric field of the light is indicated by the color (blue, orange, and purple for D$_1$, D$_2$, and C$_2$ respectively), and the orientation of the magnetic field of the light by the marker (square, triangle, and circle for D$_1$, D$_2$, and C$_2$ respectively). Filled (unfilled) markers indicate a detection polarization parallel to (perpendicular to) the input polarization. The only signals displaying a phase shift of the modulation are (a) and (b), of \SI{9\pm 1}{\degree} and \SI{-8 \pm 1}{\degree} for parallel and perpendicular signals respectively.   }
\end{figure*}
\begin{figure*}
	\centering
    \includegraphics[width=2.0\columnwidth]{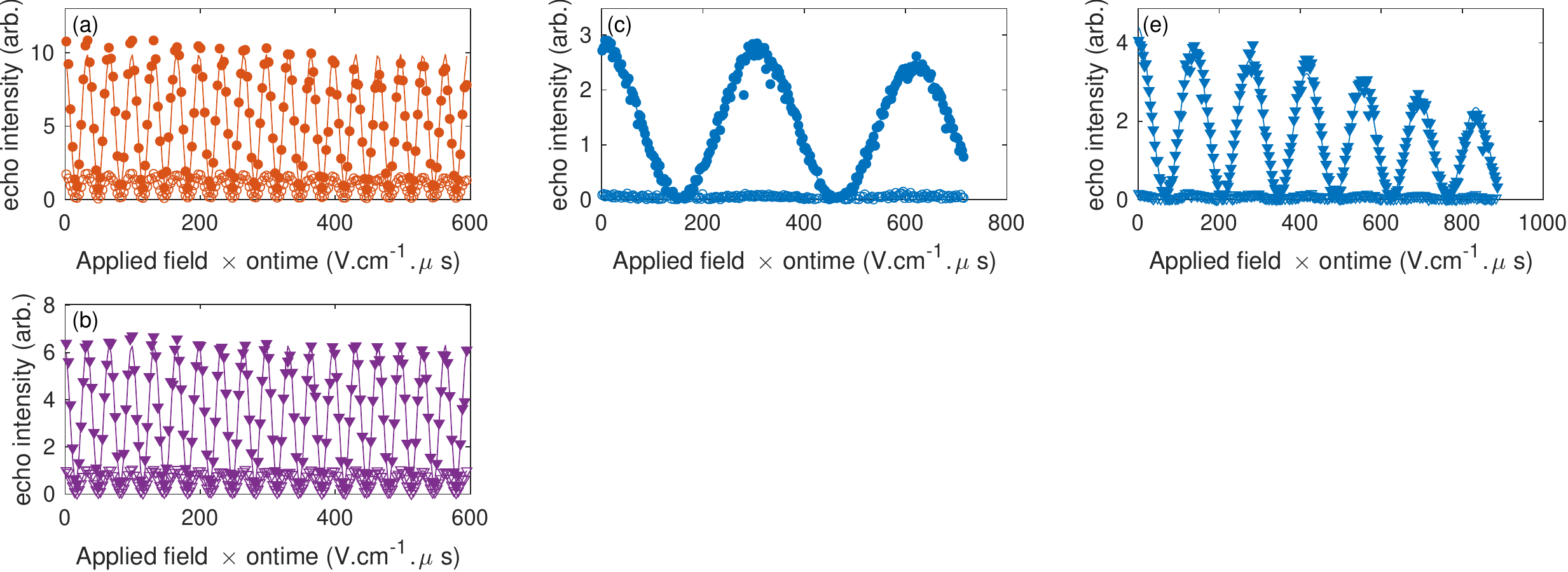}
    \caption{\label{S2mod} Stark echo modulations for site 2 \ert:\yso for three different combinations of propagation and applied electric field  directions, for two different input polarizations (linear along optical extinction axes) and two different detected polarizations (linear along optical extinction axes). (a) and (b), propagation along D$_1$, electric field along C$_2$. (c)  propagation along D$_2$, electric field along D$_1$. (e) , propagation along C$_2$, electric field along D$_2$. For D$_2$ and C$_2$ propagation, only one input polarization is presented as no echo is seen in the orthogonal polarization. In all sub-figures, the orientation of the electric field of the light is indicated by the color (blue, orange, and purple for D$_1$, D$_2$, and C$_2$ respectively), and the orientation of the magnetic field of the light by the marker (square, triangle, and circle for D$_1$, D$_2$, and C$_2$ respectively). Filled (unfilled) markers indicate a detection polarization parallel to (perpendicular to) the input polarization.    }
\end{figure*}

Examples of the integrated intensity of the Stark modulated echo are shown for site 1 and site 2 in Figs. \ref{S1mod} and \ref{S2mod} respectively. These data were taken on the larger (lower Zeeman) peak in the spectrum. The selection of figures samples all combinations of propagation, polarization, and applied electric field along the D$_1$, D$_2$, and C$_2$ axes for which echoes were seen.  

Echoes were seen for nearly all combinations of directions and polarization except for site 2 with propagation along D$_2$ and C$_2$ and polarization perpendicular to D$_1$.  We may have seen echoes in these directions had we used longer excitation pulses, but we kept $\pi$ pulses fixed below \SI{1}{\micro\second}  for all polarizations with the same propagation direction to reduce noise due to laser frequency instabilities.

The figures display all three indicators of a mixed moment: reduced visibility and phase shifts of the modulation and polarization rotation of the signal. Both the phase shifts and polarization rotation are only significant for propagation along the C$_2$ direction, and even in that direction are small. The reduction of visibility is more significant, especially for site 1, reaching as low as 55\% in the C$_2$ direction. Site 2 shows a reduction in visibility along the C$_2$ direction only, to 93\%. These data suggest that site 1 has substantial electric and magnetic dipole components in most directions, while site 2 is dominated by one component. Previous absorption measurements taken only for propagation in the D$_1$-D$_2$ plane \cite{petit20} suggested that either site 2 has only a small electric dipole component, or that the electric dipole is isotropic in the plane measured there. The results here confirm the former possibility: site 2 is predominantly magnetic dipole.

The Stark shift of \ert:\yso can also be extracted from these data by fitting the modulation frequency.  To extract the Stark shift $\Delta_s$, the modulated signal as a function of field on-time $t_{on}$ was fit with the function
\begin{equation}
    I(t_{on}) = A\left[\cos^2\left(2\pi\Delta_s \frac{V}{d} t_{on}+\phi\right) +\frac{1-W}{2W}\right]e^{-\frac{t_{on}^2}{C}}
\end{equation}
where $A$ describes the signal amplitude, $V$ is the applied voltage, $d$ is the thickness of the crystal, $\phi$ is the signal phase shift, $W$ is the signal visibility, and $C$ is a coefficient describing the small decay in the modulation seen in some spectra. This decay arises from inhomogeneity in the applied electric field, likely due to small misalignments of the capacitor plates. We chose a Gaussian decay shape as the simplest form that fits the data well enough to accurately extract the Stark shift.


\begin{figure}
	\centering
\includegraphics[width=1.0\columnwidth]{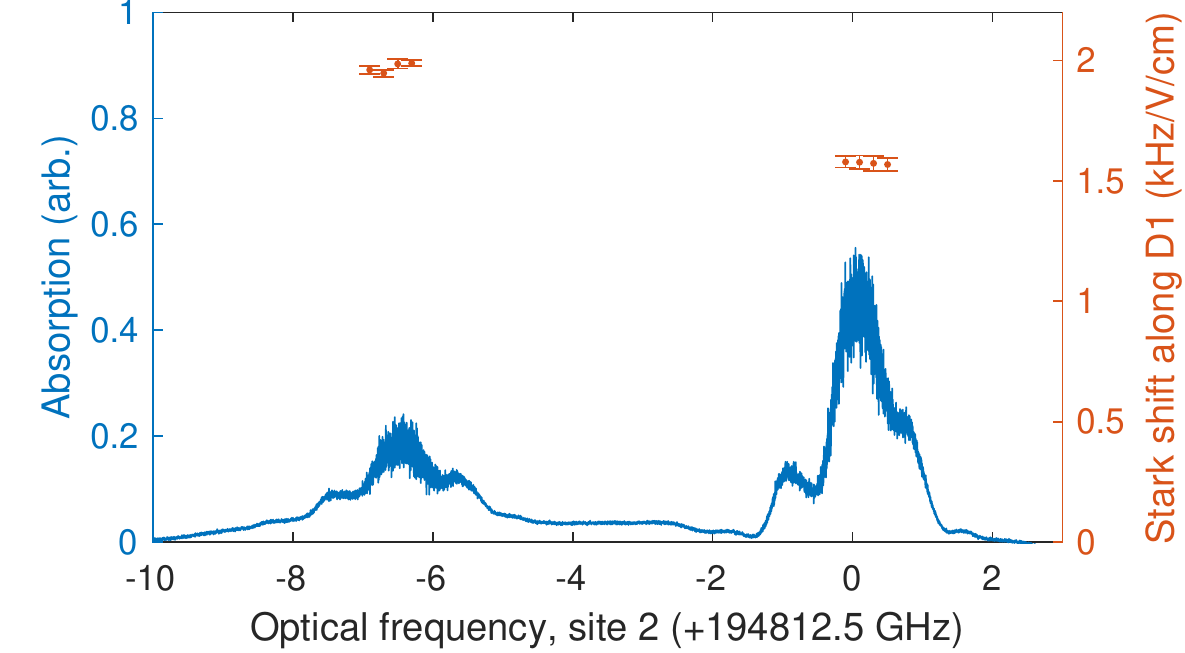}
    \caption{\label{S2stark} Variation of the Stark shift for site 2  and an applied electric field along D$_1$ (orange dots, right axis) at different frequencies in the optical spectrum (blue line, left axis). A pronounced difference in the Stark shift is evident for excitations out of the two different Zeeman sub-levels (lower sub-level on the right, upper on the left). The noise on the absorption signal is due to spectral holeburning.}
\end{figure}
While Figs. \ref{S1mod} and \ref{S2mod} only present the Stark modulated signals for the lower Zeeman transition, we measured Stark shifts across the upper and lower Zeeman peaks in the optical spectrum. We saw no variation in Stark shift for different frequencies within one peak (corresponding to driving transitions between different hyperfine states) but in some directions, we did see a substantial difference in the Stark shift of upper and lower Zeeman transitions. The largest difference (30\%) was for site 2 with a field in the D$_2$ direction, shown in Fig. \ref{S2stark}.

 \begin{table}
    \centering
\begin{ruledtabular}\begin{tabular}{lccc}
	&		&	\multicolumn{2}{c}{Stark shift (\SI{}{\kilo\hertz\per(\volt\per\cm)})}			\\
	&	Field direction	&	Lower Zeeman	&	Upper Zeeman	\\
 \hline
Site 1	&	D$_1$	&	\SI{11.93 \pm 0.05}{}	&	--	\\
	&	D$_2$	&	\SI{11.07\pm 0.04}{}	&	\SI{11.54\pm 0.04}{}	\\
	&	C$_2$	&	\SI{10.50\pm 0.03}{}	&	--	\\
\hline
Site 2	&	D$_1$	&	\SI{1.61\pm 0.01}{}	&	\SI{2.12\pm 0.01}{}	\\
	&	D$_2$	&	\SI{3.59\pm 0.01}{}	&	\SI{3.50\pm 0.01}{}	\\
	&	C$_2$	&	\SI{15.35\pm 0.05}{}	&	\SI{15.49\pm 0.05}{}	
\end{tabular}
    \end{ruledtabular}
    \caption{Measured Stark shifts for the \ertrans Z$_1\rightarrow$Y$_1$ transition of \ert:\yso. Where a value is missing for the upper Zeeman transition, that Stark shift is identical to the lower Zeeman transition. Uncertainties are only for the fit and do not include uncertainties arising from misalignment of the electric field, caused, for instance, by inaccuracy in the orientation of the cut surfaces of the crystal, estimated to be $<1^\circ$. }
    \label{stark}
\end{table}
  All the extracted Stark shifts are given in Table \ref{stark} for both the upper Zeeman and the lower Zeeman transition. In cases where these were identical, a result is only presented for the lower Zeeman transition. In determining these values, we did check that any misalignment of the field (indicated by a modulation decay) was not sufficient to reduce the electric field by comparing the fitted Stark shift to those obtained in earlier measurements we made which did not show a decay (these measurements were not presented here because they did not probe all combinations of propagation, polarization, and field directions). Only for the D$_1$ propagation, C$_2$ electric field data (Figures \ref{S1mod} and \ref{S2mod} (a) and (b))  did we observe a smaller shift (by 2\%), so the Stark shift for this direction in Table \ref{stark} derives from that earlier data. These earlier data were  taken using the same sample holder as the C$_2$ propagation data, which better constrained the alignment of the capacitor plates.

\begin{figure}
	\centering
\includegraphics[width=1.0\columnwidth]{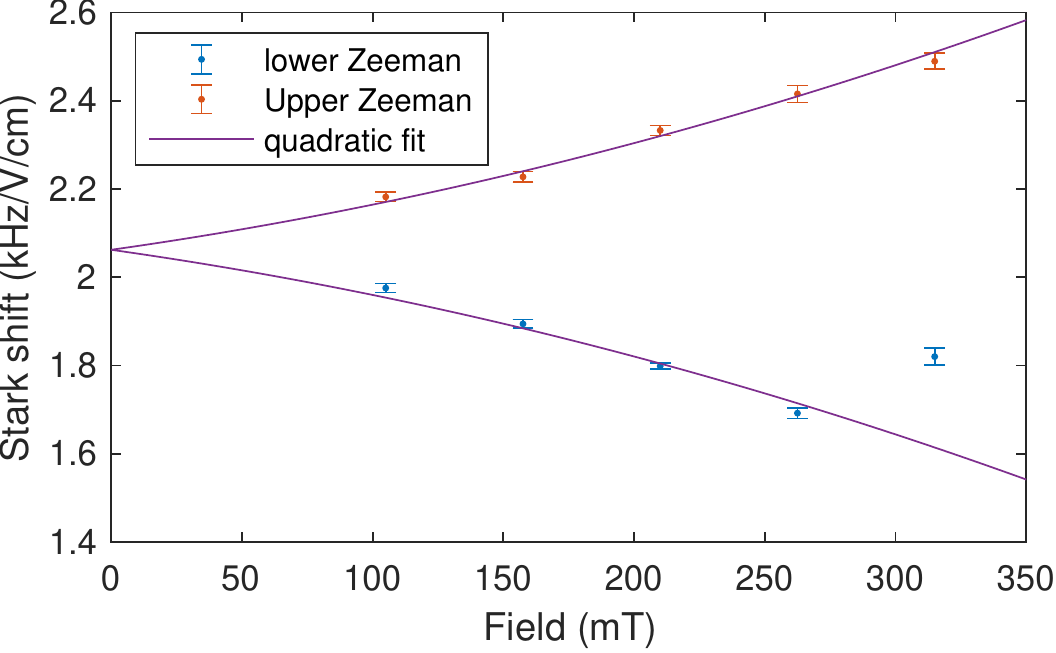}
    \caption{\label{S2starkfield} Variation of the Stark shift for the upper and lower Zeeman peaks of site 2  and an applied electric field along D$_1$ as a function of magnetic field along D$_1$. The lines are a quadratic fit to both peaks, ignoring the 315~mT lower Zeeman point for reasons discussed in the text.}
\end{figure}
To further investigate the different Stark shifts on upper and lower Zeeman peaks, we measured the Stark shift for site 2 in the D$_1$ direction as a function of magnetic field. The crystal was mounted in a home-built superconducting vector magnet with the D$_1$ direction along the $X$ direction of the magnet. Two 8~mm diameter brass bolts (which were a loose fit into the $X$-axis bore) were used as capacitor plates, and spring-loaded to ensure good electrical contact with the crystal. The light propagated along the C$_2$ axis and was polarised along D$_1$. We supplied the $X$-coil with currents up to 18~A, providing up to 350~mT field along the $D_1$ direction. We observe a change in the Stark shift that was well-fitted by a quadratic dependence on the magnetic field. The discrepancy between the fit and the highest field data point for the lower Zeeman transition in Fig. \ref{S2starkfield} could be due to a failure of one of the current power supplies used, since the earlier permanent magnet gave a Stark shift of \SI{1.61}{\kilo\hertz\per(\volt\per\cm)}, in agreement with the fit, at a similar field.

\section{Discussion}
Stark shifts in the rare earth ions arise from the mixing of opposite parity states into the $4f$ configuration by odd-parity crystal field terms \cite{bloembergen61}. The resulting sensitivity to electric fields can be seen on optical \cite{macfarlane07}, electron spin \cite{ludwig61,ham61}, and nuclear spin transitions (through the hyperfine \cite{bloembergen61a} and nuclear quadrupole \cite{armstrong61, kushida61} interactions) in different circumstances. In our measurements on the \ertrans  optical transition, we are sensitive to the relative shift of the optical ground and excited states due to their acquired electric dipole moment, and any changes in the relative electronic Zeeman splitting  due to mixing into the ground or excited state by the electric field of other crystal field components of the corresponding multiplet, known as ``$g$-shifts" \cite{kiel66}. The dominant term is the optical shift, here of the order of \SI{10}{\kilo\hertz\per(\volt\per\cm)}. This is relatively small compared to other rare earth transitions \cite{macfarlane07} because we are only sensitive to the difference in \gstate Z$_1$ and \estate Y$_1$ moments, and the odd parity mixing into these terms, which share $L$ and $S$, is similar \cite{macfarlane07}. The shifts seen here are comparable to those measured on this same transition in Er:LiNbO$_3$ and silicate glass \cite{hastings-simon06}: (\SI{15}{\kilo\hertz\per(\volt\per\cm)} and \SI{25}{\kilo\hertz\per(\volt\per\cm)}, respectively. 

 The only optical Stark shifts previously reported for the \ertrans transition of \ert:\yso are for the lower Zeeman transition of site 2 along the D$_2$ and C$_2$ axes, values of \SI{3.0\pm0.3}{\kilo\hertz\per(\volt\per\cm)} and \SI{15.0\pm0.9}{\kilo\hertz\per(\volt\per\cm)} respectively \cite{craiciu20}. The C$_2$ value is in very good agreement with our value of \SI{15.35\pm0.05}{\kilo\hertz\per(\volt\per\cm)}. The small discrepancy with our D$_2$ value and \SI{3.59\pm0.01}{\kilo\hertz\per(\volt\per\cm)} may be due to a slight misalignment of the electric field between the two measurements, or associated with the different magnetic fields used (1~T compared to our 300~mT).

 We see clear evidence of $g$-shifts, appearing as different Stark shifts measured for the lower and upper Zeeman transitions: the Stark shift of the two transitions is dependent on the relative optical Stark shift  $\Delta_s^{(o)}$ and the relative $g$-shift $\Delta_s^{(g)}$ according to $\Delta_s^{(o)} \pm \Delta_s^{(g)}$.  We saw $g$-shifts in all three directions in site 2 and for the D$_2$ direction in site 1. The $g$-shift for site 2 for electric and magnetic fields in the D$_1$ direction was dependent quadratically on the magnetic field at a rate of \SI{10}{\kilo\hertz\per\tesla^2\per(\volt\per\cm)}. This $g$-shift is only slightly smaller in size to those seen previously in ground state electron paramagnetic resonance measurements in \ert:CaWO$_4$ \cite{mims65} of between 1.5 and \SI{16}{\kilo\hertz\per\tesla\per(\volt\per\cm)} (noting that we measured the \emph{relative} shift of the ground and excited states, while those measurements were absolute shifts), but those shifts were linear in the magnetic field in accordance with the theory \cite{kiel66}. The quadratic shift here, thus, requires further investigation to confirm and better characterize the dependence in order to understand its origin. Direct measurements of the electron spin transition would be valuable to disentangle the effects of the ground and excited states.

The $g$-shift here has implications for single-instance applications of Er spins in quantum information such as for quantum light sources \cite{chen20,kindem20}. These applications require control and individual addressability of individual spins, commonly achieved using a combination of  fixed magnetic fields with frequency-selective optical or radio frequency driving. For on-chip control in a  photonic device containing several distinct spins, electric fields are preferred over magnetic fields because they are much easier to apply and localize. Electrical control of an \emph{optical} quantum state has already been demonstrated in \ert:\yso; the $g$-shift here indicates that electrical control of the spin state is also possible, using controlled phase gates as has already been demonstrated in Ce\tplus:YAG \cite{liu20}. Further, the $g$-shift is associated with a finite transition electric dipole moment, and, indeed, \ert:CaWO$_4$ measurements described above showed some evidence of a possible electric dipole transition \cite{mims65}. This could allow the electric dipole spin resonance techniques applied to other single spin qubits \cite{corna18,krauth22,osika22} to be used in \ert. Finally, the linear sensitivity to electric fields does indicate that the spin transition is sensitive to electrical noise. Full characterization of the $g$-shift of the ground state would be useful to determine this sensitivity in applications where  the spin is exposed to electric noise, such as in near-surface ions.

The optical transition dipole moment of the \ertrans Z$_1\rightarrow$Y$_1$ transition in \ert:\yso  is only partly characterized in the literature: Sun \cite{sun05} reported absorption coefficients for light polarised in all three principal axes directions without specifying the propagation direction (which is needed to interpret a mixed dipole moment), and Petit et al. \cite{petit20} measured absorption coefficients for light propagating in the D$_1$-D$_2$ plane. Combined with Petit et al.'s measurements, our results show that site 2 has primarily a magnetic dipole transition and that site 1 is strongly mixed, particularly for D$_2$ polarization, where the low visibility seen here suggests that the transition is near equal parts electric and magnetic dipole.


The hybrid dipole moment of site 1 has consequences for quantum memory applications of this crystal. As discussed in Section \ref{OBE}, when the electric and magnetic transition dipole contributions are similar, the optical Rabi frequencies for the two inversion-related sub-sites will be quite different. It may become difficult to drive both subsets identically with optical pulses, even those designed to be robust to variations in the optical coupling, such as complex hyberbolic secant (CHS) pulses. This effect may explain why, in demonstrations of the ROSE protocol \cite{damon11} in site 1 \ert:\yso, the efficiency of CHS $\pi$ pulses was only \SI{\approx 80}{\percent} \cite{dajczgewand14}. Those measurements were made with light polarised along D$_2$, the most mixed direction seen here. In contrast, in recent quantum memory experiments using site 2 \ert, we have seen pulse efficiency of 92\% using simple Gaussian $\pi$ pulses. Measurements made on a single sub-site (obtained using holeburning in an applied electric field) could confirm if the low-efficiency site 1 is due to different sub-site Rabi frequencies.

While the different Rabi frequencies of the sub-sites can present a challenge for quantum memory applications, the hybrid moment presents several opportunities as well. Fig. \ref{modexamples} shows that the applied electric field can shift the polarization of the output signal (dependent on the shape of the electric and magnetic dipole moments). We did not see large polarization shifts in the experimental data, but this is not surprising: due to the birefringence, there is a phase-matching condition that suppresses the echo in the polarization orthogonal to the input. It is likely this phase-matching condition can be modified to allow orthogonal output. Then, an electric field could be used to alter the polarization of a quantum state, allowing linear optical processing of the quantum state. 

These quantum memory applications would, however, first require a full characterization of the electric and magnetic dipole moments and of the propagation of light through the crystal to identify the appropriate conditions. The required quantitative information about the transition dipole moment can be determined by fitting a Maxwell-Bloch model to Stark-modulated photon echo measurements. We cannot use the simple  Optical Bloch model here because it does not account for birefringence. Even though we propagated along principal axes of the refractive index tensor (so that the input light polarization did not vary across the crystal), the birefringence is still important because the atoms radiate according to the orientations of the electric and magnetic dipole moments, and these are not constrained to coincide with the refractive index tensor. While the absorption behaviour of light propagation in absorbing materials is well understood, and has, indeed, been studied in \yso in the visible \cite{sabooni16}, a full Maxwell Bloch model for the coherent interaction has not yet been made. Implementation of such a model for \ert:\yso is hampered by the lack of information about the refractive index tensor at 1540~nm. Refractive indices have been measured at visible wavelengths \cite{beach90a} but extrapolation of the fitted curves to 1540~nm produces unphysical behavior. We will present a full Maxwell Bloch model, including measurements of the refractive index, in subsequent work.

\section{Conclusion}
We have measured the Stark shift of the \ertrans transition of \ert:\yso Z$_1\rightarrow$Y$_1$ along the three principal axes of the refractive index tensor, finding a fairly isotropic, linear Stark shift for site 1 with an average value of \SI{11.2}{\kilo\hertz\per(\volt\per\cm)} and an anisotropic Stark shift for site 2 varying between approximately 2 and \SI{15}{\kilo\hertz\per(\volt\per\cm)}. For the D$_2$ direction of site 1 and for all three directions in site 2, we observed a $g$-shift component of the Stark shift in addition to the optical component, contributing up to \SI{0.5}{\kilo\hertz\per(\volt\per\cm)} at our applied magnetic field of 300~mT.

\bibliographystyle{apsrev4-2}
\bibliography{mainbib}

\begin{thebibliography}{44}%
\makeatletter
\providecommand \@ifxundefined [1]{%
 \@ifx{#1\undefined}
}%
\providecommand \@ifnum [1]{%
 \ifnum #1\expandafter \@firstoftwo
 \else \expandafter \@secondoftwo
 \fi
}%
\providecommand \@ifx [1]{%
 \ifx #1\expandafter \@firstoftwo
 \else \expandafter \@secondoftwo
 \fi
}%
\providecommand \natexlab [1]{#1}%
\providecommand \enquote  [1]{``#1''}%
\providecommand \bibnamefont  [1]{#1}%
\providecommand \bibfnamefont [1]{#1}%
\providecommand \citenamefont [1]{#1}%
\providecommand \href@noop [0]{\@secondoftwo}%
\providecommand \href [0]{\begingroup \@sanitize@url \@href}%
\providecommand \@href[1]{\@@startlink{#1}\@@href}%
\providecommand \@@href[1]{\endgroup#1\@@endlink}%
\providecommand \@sanitize@url [0]{\catcode `\\12\catcode `\$12\catcode
  `\&12\catcode `\#12\catcode `\^12\catcode `\_12\catcode `\%12\relax}%
\providecommand \@@startlink[1]{}%
\providecommand \@@endlink[0]{}%
\providecommand \url  [0]{\begingroup\@sanitize@url \@url }%
\providecommand \@url [1]{\endgroup\@href {#1}{\urlprefix }}%
\providecommand \urlprefix  [0]{URL }%
\providecommand \Eprint [0]{\href }%
\providecommand \doibase [0]{https://doi.org/}%
\providecommand \selectlanguage [0]{\@gobble}%
\providecommand \bibinfo  [0]{\@secondoftwo}%
\providecommand \bibfield  [0]{\@secondoftwo}%
\providecommand \translation [1]{[#1]}%
\providecommand \BibitemOpen [0]{}%
\providecommand \bibitemStop [0]{}%
\providecommand \bibitemNoStop [0]{.\EOS\space}%
\providecommand \EOS [0]{\spacefactor3000\relax}%
\providecommand \BibitemShut  [1]{\csname bibitem#1\endcsname}%
\let\auto@bib@innerbib\@empty
\bibitem [{\citenamefont {Rochman}\ \emph {et~al.}(2023)\citenamefont
  {Rochman}, \citenamefont {Xie}, \citenamefont {Bartholomew}, \citenamefont
  {Schwab},\ and\ \citenamefont {Faraon}}]{rochman23}%
  \BibitemOpen
  \bibfield  {author} {\bibinfo {author} {\bibfnamefont {J.}~\bibnamefont
  {Rochman}}, \bibinfo {author} {\bibfnamefont {T.}~\bibnamefont {Xie}},
  \bibinfo {author} {\bibfnamefont {J.~G.}\ \bibnamefont {Bartholomew}},
  \bibinfo {author} {\bibfnamefont {K.~C.}\ \bibnamefont {Schwab}},\ and\
  \bibinfo {author} {\bibfnamefont {A.}~\bibnamefont {Faraon}},\ }\href
  {https://doi.org/10.1038/s41467-023-36799-0} {\bibfield  {journal} {\bibinfo
  {journal} {Nature Communications}\ }\textbf {\bibinfo {volume} {14}},\
  \bibinfo {pages} {1153} (\bibinfo {year} {2023})}\BibitemShut {NoStop}%
\bibitem [{\citenamefont {Chen}\ \emph {et~al.}(2020)\citenamefont {Chen},
  \citenamefont {Raha}, \citenamefont {Phenicie}, \citenamefont {Ourari},\ and\
  \citenamefont {Thompson}}]{chen20}%
  \BibitemOpen
  \bibfield  {author} {\bibinfo {author} {\bibfnamefont {S.}~\bibnamefont
  {Chen}}, \bibinfo {author} {\bibfnamefont {M.}~\bibnamefont {Raha}}, \bibinfo
  {author} {\bibfnamefont {C.~M.}\ \bibnamefont {Phenicie}}, \bibinfo {author}
  {\bibfnamefont {S.}~\bibnamefont {Ourari}},\ and\ \bibinfo {author}
  {\bibfnamefont {J.~D.}\ \bibnamefont {Thompson}},\ }\href
  {https://doi.org/10.1126/science.abc7821} {\bibfield  {journal} {\bibinfo
  {journal} {Science}\ }\textbf {\bibinfo {volume} {370}},\ \bibinfo {pages}
  {592} (\bibinfo {year} {2020})}\BibitemShut {NoStop}%
\bibitem [{\citenamefont {Gritsch}\ \emph {et~al.}(2022)\citenamefont
  {Gritsch}, \citenamefont {Weiss}, \citenamefont {Fr{\"u}h}, \citenamefont
  {Rinner},\ and\ \citenamefont {Reiserer}}]{gritsch22}%
  \BibitemOpen
  \bibfield  {author} {\bibinfo {author} {\bibfnamefont {A.}~\bibnamefont
  {Gritsch}}, \bibinfo {author} {\bibfnamefont {L.}~\bibnamefont {Weiss}},
  \bibinfo {author} {\bibfnamefont {J.}~\bibnamefont {Fr{\"u}h}}, \bibinfo
  {author} {\bibfnamefont {S.}~\bibnamefont {Rinner}},\ and\ \bibinfo {author}
  {\bibfnamefont {A.}~\bibnamefont {Reiserer}},\ }\href
  {https://doi.org/10.1103/PhysRevX.12.041009} {\bibfield  {journal} {\bibinfo
  {journal} {Physical Review X}\ }\textbf {\bibinfo {volume} {12}},\ \bibinfo
  {pages} {041009} (\bibinfo {year} {2022})}\BibitemShut {NoStop}%
\bibitem [{\citenamefont {Berkman}\ \emph {et~al.}(2023)\citenamefont
  {Berkman}, \citenamefont {Lyasota}, \citenamefont {{de Boo}}, \citenamefont
  {Bartholomew}, \citenamefont {Johnson}, \citenamefont {McCallum},
  \citenamefont {Xu}, \citenamefont {Xie}, \citenamefont {Ahlefeldt},
  \citenamefont {Sellars}, \citenamefont {Yin},\ and\ \citenamefont
  {Rogge}}]{berkman23}%
  \BibitemOpen
  \bibfield  {author} {\bibinfo {author} {\bibfnamefont {I.~R.}\ \bibnamefont
  {Berkman}}, \bibinfo {author} {\bibfnamefont {A.}~\bibnamefont {Lyasota}},
  \bibinfo {author} {\bibfnamefont {G.~G.}\ \bibnamefont {{de Boo}}}, \bibinfo
  {author} {\bibfnamefont {J.~G.}\ \bibnamefont {Bartholomew}}, \bibinfo
  {author} {\bibfnamefont {B.~C.}\ \bibnamefont {Johnson}}, \bibinfo {author}
  {\bibfnamefont {J.~C.}\ \bibnamefont {McCallum}}, \bibinfo {author}
  {\bibfnamefont {B.-B.}\ \bibnamefont {Xu}}, \bibinfo {author} {\bibfnamefont
  {S.}~\bibnamefont {Xie}}, \bibinfo {author} {\bibfnamefont {R.~L.}\
  \bibnamefont {Ahlefeldt}}, \bibinfo {author} {\bibfnamefont {M.~J.}\
  \bibnamefont {Sellars}}, \bibinfo {author} {\bibfnamefont {C.}~\bibnamefont
  {Yin}},\ and\ \bibinfo {author} {\bibfnamefont {S.}~\bibnamefont {Rogge}},\
  }\href {https://doi.org/10.1103/PhysRevApplied.19.014037} {\bibfield
  {journal} {\bibinfo  {journal} {Physical Review Applied}\ }\textbf {\bibinfo
  {volume} {19}},\ \bibinfo {pages} {014037} (\bibinfo {year}
  {2023})}\BibitemShut {NoStop}%
\bibitem [{\citenamefont {Stuart}\ \emph {et~al.}(2021)\citenamefont {Stuart},
  \citenamefont {Hedges}, \citenamefont {Ahlefeldt},\ and\ \citenamefont
  {Sellars}}]{stuart21}%
  \BibitemOpen
  \bibfield  {author} {\bibinfo {author} {\bibfnamefont {J.~S.}\ \bibnamefont
  {Stuart}}, \bibinfo {author} {\bibfnamefont {M.}~\bibnamefont {Hedges}},
  \bibinfo {author} {\bibfnamefont {R.}~\bibnamefont {Ahlefeldt}},\ and\
  \bibinfo {author} {\bibfnamefont {M.}~\bibnamefont {Sellars}},\ }\href
  {https://doi.org/10.1103/PhysRevResearch.3.L032054} {\bibfield  {journal}
  {\bibinfo  {journal} {Physical Review Research}\ }\textbf {\bibinfo {volume}
  {3}},\ \bibinfo {pages} {L032054} (\bibinfo {year} {2021})}\BibitemShut
  {NoStop}%
\bibitem [{\citenamefont {Weber}(1967)}]{weber67}%
  \BibitemOpen
  \bibfield  {author} {\bibinfo {author} {\bibfnamefont {M.~J.}\ \bibnamefont
  {Weber}},\ }\href {https://doi.org/10.1103/PhysRev.157.262} {\bibfield
  {journal} {\bibinfo  {journal} {Physical Review}\ }\textbf {\bibinfo {volume}
  {157}},\ \bibinfo {pages} {262} (\bibinfo {year} {1967})}\BibitemShut
  {NoStop}%
\bibitem [{\citenamefont {Weber}(1968)}]{weber68}%
  \BibitemOpen
  \bibfield  {author} {\bibinfo {author} {\bibfnamefont {M.~J.}\ \bibnamefont
  {Weber}},\ }\href {https://doi.org/10.1103/PhysRev.171.283} {\bibfield
  {journal} {\bibinfo  {journal} {Physical Review}\ }\textbf {\bibinfo {volume}
  {171}},\ \bibinfo {pages} {283} (\bibinfo {year} {1968})}\BibitemShut
  {NoStop}%
\bibitem [{\citenamefont {Weber}\ \emph {et~al.}(1973)\citenamefont {Weber},
  \citenamefont {Varitimos},\ and\ \citenamefont {Matsinger}}]{weber73}%
  \BibitemOpen
  \bibfield  {author} {\bibinfo {author} {\bibfnamefont {M.~J.}\ \bibnamefont
  {Weber}}, \bibinfo {author} {\bibfnamefont {T.~E.}\ \bibnamefont
  {Varitimos}},\ and\ \bibinfo {author} {\bibfnamefont {B.~H.}\ \bibnamefont
  {Matsinger}},\ }\href {https://doi.org/10.1103/PhysRevB.8.47} {\bibfield
  {journal} {\bibinfo  {journal} {Physical Review B}\ }\textbf {\bibinfo
  {volume} {8}},\ \bibinfo {pages} {47} (\bibinfo {year} {1973})}\BibitemShut
  {NoStop}%
\bibitem [{\citenamefont {Li}\ \emph {et~al.}(2014)\citenamefont {Li},
  \citenamefont {Jiang}, \citenamefont {Cueff}, \citenamefont {Dodson},
  \citenamefont {Karaveli},\ and\ \citenamefont {Zia}}]{li14c}%
  \BibitemOpen
  \bibfield  {author} {\bibinfo {author} {\bibfnamefont {D.}~\bibnamefont
  {Li}}, \bibinfo {author} {\bibfnamefont {M.}~\bibnamefont {Jiang}}, \bibinfo
  {author} {\bibfnamefont {S.}~\bibnamefont {Cueff}}, \bibinfo {author}
  {\bibfnamefont {C.~M.}\ \bibnamefont {Dodson}}, \bibinfo {author}
  {\bibfnamefont {S.}~\bibnamefont {Karaveli}},\ and\ \bibinfo {author}
  {\bibfnamefont {R.}~\bibnamefont {Zia}},\ }\href
  {https://doi.org/10.1103/PhysRevB.89.161409} {\bibfield  {journal} {\bibinfo
  {journal} {Physical Review B}\ }\textbf {\bibinfo {volume} {89}},\ \bibinfo
  {pages} {161409} (\bibinfo {year} {2014})}\BibitemShut {NoStop}%
\bibitem [{\citenamefont {Gerasimov}\ \emph {et~al.}(2016)\citenamefont
  {Gerasimov}, \citenamefont {Minnegaliev}, \citenamefont {Malkin},
  \citenamefont {Baibekov},\ and\ \citenamefont {Moiseev}}]{gerasimov16}%
  \BibitemOpen
  \bibfield  {author} {\bibinfo {author} {\bibfnamefont {K.~I.}\ \bibnamefont
  {Gerasimov}}, \bibinfo {author} {\bibfnamefont {M.~M.}\ \bibnamefont
  {Minnegaliev}}, \bibinfo {author} {\bibfnamefont {B.~Z.}\ \bibnamefont
  {Malkin}}, \bibinfo {author} {\bibfnamefont {E.~I.}\ \bibnamefont
  {Baibekov}},\ and\ \bibinfo {author} {\bibfnamefont {S.~A.}\ \bibnamefont
  {Moiseev}},\ }\href {https://doi.org/10.1103/PhysRevB.94.054429} {\bibfield
  {journal} {\bibinfo  {journal} {Physical Review B}\ }\textbf {\bibinfo
  {volume} {94}},\ \bibinfo {pages} {054429} (\bibinfo {year}
  {2016})}\BibitemShut {NoStop}%
\bibitem [{\citenamefont {Xie}\ \emph {et~al.}(2021)\citenamefont {Xie},
  \citenamefont {Rochman}, \citenamefont {Bartholomew}, \citenamefont {Ruskuc},
  \citenamefont {Kindem}, \citenamefont {Craiciu}, \citenamefont {Thiel},
  \citenamefont {Cone},\ and\ \citenamefont {Faraon}}]{Xie21}%
  \BibitemOpen
  \bibfield  {author} {\bibinfo {author} {\bibfnamefont {T.}~\bibnamefont
  {Xie}}, \bibinfo {author} {\bibfnamefont {J.}~\bibnamefont {Rochman}},
  \bibinfo {author} {\bibfnamefont {J.~G.}\ \bibnamefont {Bartholomew}},
  \bibinfo {author} {\bibfnamefont {A.}~\bibnamefont {Ruskuc}}, \bibinfo
  {author} {\bibfnamefont {J.~M.}\ \bibnamefont {Kindem}}, \bibinfo {author}
  {\bibfnamefont {I.}~\bibnamefont {Craiciu}}, \bibinfo {author} {\bibfnamefont
  {C.~W.}\ \bibnamefont {Thiel}}, \bibinfo {author} {\bibfnamefont {R.~L.}\
  \bibnamefont {Cone}},\ and\ \bibinfo {author} {\bibfnamefont
  {A.}~\bibnamefont {Faraon}},\ }\href
  {https://doi.org/10.1103/PhysRevB.104.054111} {\bibfield  {journal} {\bibinfo
   {journal} {Physical Review B}\ }\textbf {\bibinfo {volume} {104}},\ \bibinfo
  {pages} {054111} (\bibinfo {year} {2021})}\BibitemShut {NoStop}%
\bibitem [{\citenamefont {Wang}\ and\ \citenamefont {Meltzer}(1992)}]{wang92}%
  \BibitemOpen
  \bibfield  {author} {\bibinfo {author} {\bibfnamefont {Y.~P.}\ \bibnamefont
  {Wang}}\ and\ \bibinfo {author} {\bibfnamefont {R.~S.}\ \bibnamefont
  {Meltzer}},\ }\href {https://doi.org/10.1103/PhysRevB.45.10119} {\bibfield
  {journal} {\bibinfo  {journal} {Physical Review B}\ }\textbf {\bibinfo
  {volume} {45}},\ \bibinfo {pages} {10119} (\bibinfo {year}
  {1992})}\BibitemShut {NoStop}%
\bibitem [{\citenamefont {Meixner}\ \emph {et~al.}(1992)\citenamefont
  {Meixner}, \citenamefont {Jefferson},\ and\ \citenamefont
  {Macfarlane}}]{meixner92}%
  \BibitemOpen
  \bibfield  {author} {\bibinfo {author} {\bibfnamefont {A.~J.}\ \bibnamefont
  {Meixner}}, \bibinfo {author} {\bibfnamefont {C.~M.}\ \bibnamefont
  {Jefferson}},\ and\ \bibinfo {author} {\bibfnamefont {R.~M.}\ \bibnamefont
  {Macfarlane}},\ }\href {https://doi.org/10.1103/PhysRevB.46.5912} {\bibfield
  {journal} {\bibinfo  {journal} {Physical Review B}\ }\textbf {\bibinfo
  {volume} {46}},\ \bibinfo {pages} {5912} (\bibinfo {year}
  {1992})}\BibitemShut {NoStop}%
\bibitem [{\citenamefont {Macfarlane}\ and\ \citenamefont
  {Meixner}(1994)}]{macfarlane94a}%
  \BibitemOpen
  \bibfield  {author} {\bibinfo {author} {\bibfnamefont {R.~M.}\ \bibnamefont
  {Macfarlane}}\ and\ \bibinfo {author} {\bibfnamefont {A.~J.}\ \bibnamefont
  {Meixner}},\ }\href {https://doi.org/10.1364/OL.19.000987} {\bibfield
  {journal} {\bibinfo  {journal} {Optics Letters}\ }\textbf {\bibinfo {volume}
  {19}},\ \bibinfo {pages} {987} (\bibinfo {year} {1994})}\BibitemShut
  {NoStop}%
\bibitem [{\citenamefont {Graf}\ \emph {et~al.}(1997)\citenamefont {Graf},
  \citenamefont {Renn}, \citenamefont {Wild},\ and\ \citenamefont
  {Mitsunaga}}]{graf97}%
  \BibitemOpen
  \bibfield  {author} {\bibinfo {author} {\bibfnamefont {F.~R.}\ \bibnamefont
  {Graf}}, \bibinfo {author} {\bibfnamefont {A.}~\bibnamefont {Renn}}, \bibinfo
  {author} {\bibfnamefont {U.~P.}\ \bibnamefont {Wild}},\ and\ \bibinfo
  {author} {\bibfnamefont {M.}~\bibnamefont {Mitsunaga}},\ }\href
  {https://doi.org/10.1103/PhysRevB.55.11225} {\bibfield  {journal} {\bibinfo
  {journal} {Physical Review B}\ }\textbf {\bibinfo {volume} {55}},\ \bibinfo
  {pages} {11225} (\bibinfo {year} {1997})}\BibitemShut {NoStop}%
\bibitem [{\citenamefont {Craiciu}\ \emph {et~al.}(2021)\citenamefont
  {Craiciu}, \citenamefont {Lei}, \citenamefont {Rochman}, \citenamefont
  {Bartholomew},\ and\ \citenamefont {Faraon}}]{craiciu21}%
  \BibitemOpen
  \bibfield  {author} {\bibinfo {author} {\bibfnamefont {I.}~\bibnamefont
  {Craiciu}}, \bibinfo {author} {\bibfnamefont {M.}~\bibnamefont {Lei}},
  \bibinfo {author} {\bibfnamefont {J.}~\bibnamefont {Rochman}}, \bibinfo
  {author} {\bibfnamefont {J.~G.}\ \bibnamefont {Bartholomew}},\ and\ \bibinfo
  {author} {\bibfnamefont {A.}~\bibnamefont {Faraon}},\ }\href
  {https://doi.org/10.1364/OPTICA.412211} {\bibfield  {journal} {\bibinfo
  {journal} {Optica}\ }\textbf {\bibinfo {volume} {8}},\ \bibinfo {pages} {114}
  (\bibinfo {year} {2021})}\BibitemShut {NoStop}%
\bibitem [{\citenamefont {Min{\'a}{\v r}}\ \emph {et~al.}(2009)\citenamefont
  {Min{\'a}{\v r}}, \citenamefont {Lauritzen}, \citenamefont {de~Riedmatten},
  \citenamefont {Afzelius}, \citenamefont {Simon},\ and\ \citenamefont
  {Gisin}}]{minar09}%
  \BibitemOpen
  \bibfield  {author} {\bibinfo {author} {\bibfnamefont {J.}~\bibnamefont
  {Min{\'a}{\v r}}}, \bibinfo {author} {\bibfnamefont {B.}~\bibnamefont
  {Lauritzen}}, \bibinfo {author} {\bibfnamefont {H.}~\bibnamefont
  {de~Riedmatten}}, \bibinfo {author} {\bibfnamefont {M.}~\bibnamefont
  {Afzelius}}, \bibinfo {author} {\bibfnamefont {C.}~\bibnamefont {Simon}},\
  and\ \bibinfo {author} {\bibfnamefont {N.}~\bibnamefont {Gisin}},\ }\href
  {https://doi.org/10.1088/1367-2630/11/11/113019} {\bibfield  {journal}
  {\bibinfo  {journal} {New Journal of Physics}\ }\textbf {\bibinfo {volume}
  {11}},\ \bibinfo {pages} {113019} (\bibinfo {year} {2009})}\BibitemShut
  {NoStop}%
\bibitem [{\citenamefont {Hetet}\ \emph {et~al.}(2008)\citenamefont {Hetet},
  \citenamefont {Longdell}, \citenamefont {Alexander}, \citenamefont {Lam},\
  and\ \citenamefont {Sellars}}]{hetet08}%
  \BibitemOpen
  \bibfield  {author} {\bibinfo {author} {\bibfnamefont {G.}~\bibnamefont
  {Hetet}}, \bibinfo {author} {\bibfnamefont {J.~J.}\ \bibnamefont {Longdell}},
  \bibinfo {author} {\bibfnamefont {A.~L.}\ \bibnamefont {Alexander}}, \bibinfo
  {author} {\bibfnamefont {P.~K.}\ \bibnamefont {Lam}},\ and\ \bibinfo {author}
  {\bibfnamefont {M.~J.}\ \bibnamefont {Sellars}},\ }\href
  {https://doi.org/10.1103/PhysRevLett.100.023601} {\bibfield  {journal}
  {\bibinfo  {journal} {Physical Review Letters}\ }\textbf {\bibinfo {volume}
  {100}},\ \bibinfo {pages} {023601} (\bibinfo {year} {2008})}\BibitemShut
  {NoStop}%
\bibitem [{\citenamefont {Damon}\ \emph {et~al.}(2011)\citenamefont {Damon},
  \citenamefont {Bonarota}, \citenamefont {{Louchet-Chauvet}}, \citenamefont
  {Chaneli{\`e}re},\ and\ \citenamefont {Le~Gou{\"e}t}}]{damon11}%
  \BibitemOpen
  \bibfield  {author} {\bibinfo {author} {\bibfnamefont {V.}~\bibnamefont
  {Damon}}, \bibinfo {author} {\bibfnamefont {M.}~\bibnamefont {Bonarota}},
  \bibinfo {author} {\bibfnamefont {A.}~\bibnamefont {{Louchet-Chauvet}}},
  \bibinfo {author} {\bibfnamefont {T.}~\bibnamefont {Chaneli{\`e}re}},\ and\
  \bibinfo {author} {\bibfnamefont {J.-L.}\ \bibnamefont {Le~Gou{\"e}t}},\
  }\href {https://doi.org/10.1088/1367-2630/13/9/093031} {\bibfield  {journal}
  {\bibinfo  {journal} {New Journal of Physics}\ }\textbf {\bibinfo {volume}
  {13}},\ \bibinfo {pages} {093031} (\bibinfo {year} {2011})}\BibitemShut
  {NoStop}%
\bibitem [{\citenamefont {McAuslan}\ \emph {et~al.}(2011)\citenamefont
  {McAuslan}, \citenamefont {Ledingham}, \citenamefont {Naylor}, \citenamefont
  {Beavan}, \citenamefont {Hedges}, \citenamefont {Sellars},\ and\
  \citenamefont {Longdell}}]{mcauslan11}%
  \BibitemOpen
  \bibfield  {author} {\bibinfo {author} {\bibfnamefont {D.~L.}\ \bibnamefont
  {McAuslan}}, \bibinfo {author} {\bibfnamefont {P.~M.}\ \bibnamefont
  {Ledingham}}, \bibinfo {author} {\bibfnamefont {W.~R.}\ \bibnamefont
  {Naylor}}, \bibinfo {author} {\bibfnamefont {S.~E.}\ \bibnamefont {Beavan}},
  \bibinfo {author} {\bibfnamefont {M.~P.}\ \bibnamefont {Hedges}}, \bibinfo
  {author} {\bibfnamefont {M.~J.}\ \bibnamefont {Sellars}},\ and\ \bibinfo
  {author} {\bibfnamefont {J.~J.}\ \bibnamefont {Longdell}},\ }\href
  {https://doi.org/10.1103/PhysRevA.84.022309} {\bibfield  {journal} {\bibinfo
  {journal} {Physical Review A}\ }\textbf {\bibinfo {volume} {84}},\ \bibinfo
  {pages} {022309} (\bibinfo {year} {2011})}\BibitemShut {NoStop}%
\bibitem [{\citenamefont {Beach}\ \emph {et~al.}(1990)\citenamefont {Beach},
  \citenamefont {Shinn}, \citenamefont {Davis}, \citenamefont {Solarz},\ and\
  \citenamefont {Krupke}}]{beach90a}%
  \BibitemOpen
  \bibfield  {author} {\bibinfo {author} {\bibfnamefont {R.}~\bibnamefont
  {Beach}}, \bibinfo {author} {\bibfnamefont {M.}~\bibnamefont {Shinn}},
  \bibinfo {author} {\bibfnamefont {L.}~\bibnamefont {Davis}}, \bibinfo
  {author} {\bibfnamefont {R.}~\bibnamefont {Solarz}},\ and\ \bibinfo {author}
  {\bibfnamefont {W.}~\bibnamefont {Krupke}},\ }\href
  {https://doi.org/10.1109/3.59689} {\bibfield  {journal} {\bibinfo  {journal}
  {IEEE Journal of Quantum Electronics}\ }\textbf {\bibinfo {volume} {26}},\
  \bibinfo {pages} {1405} (\bibinfo {year} {1990})}\BibitemShut {NoStop}%
\bibitem [{\citenamefont {Petit}\ \emph {et~al.}(2020)\citenamefont {Petit},
  \citenamefont {Boulanger}, \citenamefont {Debray},\ and\ \citenamefont
  {Chaneli{\`e}re}}]{petit20}%
  \BibitemOpen
  \bibfield  {author} {\bibinfo {author} {\bibfnamefont {Y.}~\bibnamefont
  {Petit}}, \bibinfo {author} {\bibfnamefont {B.}~\bibnamefont {Boulanger}},
  \bibinfo {author} {\bibfnamefont {J.}~\bibnamefont {Debray}},\ and\ \bibinfo
  {author} {\bibfnamefont {T.}~\bibnamefont {Chaneli{\`e}re}},\ }\href
  {https://doi.org/10.1016/j.omx.2020.100062} {\bibfield  {journal} {\bibinfo
  {journal} {Optical Materials: X}\ }\textbf {\bibinfo {volume} {8}},\ \bibinfo
  {pages} {100062} (\bibinfo {year} {2020})}\BibitemShut {NoStop}%
\bibitem [{\citenamefont {Sun}(2005)}]{sun05}%
  \BibitemOpen
  \bibfield  {author} {\bibinfo {author} {\bibfnamefont {Y.~C.}\ \bibnamefont
  {Sun}}\ }(\bibinfo  {publisher} {{Springer}},\ \bibinfo {address} {{Berlin,
  Heidelberg}},\ \bibinfo {year} {2005})\ pp.\ \bibinfo {pages}
  {379--429}\BibitemShut {NoStop}%
\bibitem [{\citenamefont {Levenson}\ and\ \citenamefont
  {Kano}(1989)}]{levenson89}%
  \BibitemOpen
  \bibfield  {author} {\bibinfo {author} {\bibfnamefont {M.~D.}\ \bibnamefont
  {Levenson}}\ and\ \bibinfo {author} {\bibfnamefont {S.~S.}\ \bibnamefont
  {Kano}},\ }\href@noop {} {\emph {\bibinfo {title} {Introduction to
  {{Nonlinear Laser Spectroscopy}}}}},\ \bibinfo {edition} {revised}\ ed.\
  (\bibinfo  {publisher} {{Academic Press}},\ \bibinfo {address} {{San
  Diego}},\ \bibinfo {year} {1989})\BibitemShut {NoStop}%
\bibitem [{\citenamefont {Macfarlane}(2007)}]{macfarlane07}%
  \BibitemOpen
  \bibfield  {author} {\bibinfo {author} {\bibfnamefont {R.~M.}\ \bibnamefont
  {Macfarlane}},\ }\href {https://doi.org/10.1016/j.jlumin.2006.08.012}
  {\bibfield  {journal} {\bibinfo  {journal} {Journal of Luminescence}\
  }\textbf {\bibinfo {volume} {125}},\ \bibinfo {pages} {156} (\bibinfo {year}
  {2007})}\BibitemShut {NoStop}%
\bibitem [{\citenamefont {Sun}\ \emph {et~al.}(2008)\citenamefont {Sun},
  \citenamefont {B{\"o}ttger}, \citenamefont {Thiel},\ and\ \citenamefont
  {Cone}}]{sun08}%
  \BibitemOpen
  \bibfield  {author} {\bibinfo {author} {\bibfnamefont {Y.}~\bibnamefont
  {Sun}}, \bibinfo {author} {\bibfnamefont {T.}~\bibnamefont {B{\"o}ttger}},
  \bibinfo {author} {\bibfnamefont {C.~W.}\ \bibnamefont {Thiel}},\ and\
  \bibinfo {author} {\bibfnamefont {R.~L.}\ \bibnamefont {Cone}},\ }\href
  {https://doi.org/10.1103/PhysRevB.77.085124} {\bibfield  {journal} {\bibinfo
  {journal} {Physical Review B}\ }\textbf {\bibinfo {volume} {77}},\ \bibinfo
  {pages} {085124} (\bibinfo {year} {2008})}\BibitemShut {NoStop}%
\bibitem [{\citenamefont {Fraval}\ \emph {et~al.}(2004)\citenamefont {Fraval},
  \citenamefont {Sellars},\ and\ \citenamefont {Longdell}}]{fraval04}%
  \BibitemOpen
  \bibfield  {author} {\bibinfo {author} {\bibfnamefont {E.}~\bibnamefont
  {Fraval}}, \bibinfo {author} {\bibfnamefont {M.~J.}\ \bibnamefont
  {Sellars}},\ and\ \bibinfo {author} {\bibfnamefont {J.~J.}\ \bibnamefont
  {Longdell}},\ }\href {https://doi.org/10.1103/PhysRevLett.92.077601}
  {\bibfield  {journal} {\bibinfo  {journal} {Physical Review Letters}\
  }\textbf {\bibinfo {volume} {92}},\ \bibinfo {pages} {077601} (\bibinfo
  {year} {2004})}\BibitemShut {NoStop}%
\bibitem [{\citenamefont {Bloembergen}(1961{\natexlab{a}})}]{bloembergen61}%
  \BibitemOpen
  \bibfield  {author} {\bibinfo {author} {\bibfnamefont {N.}~\bibnamefont
  {Bloembergen}},\ }\href {https://doi.org/10.1126/science.133.3461.1363}
  {\bibfield  {journal} {\bibinfo  {journal} {Science}\ }\textbf {\bibinfo
  {volume} {133}},\ \bibinfo {pages} {1363} (\bibinfo {year}
  {1961}{\natexlab{a}})}\BibitemShut {NoStop}%
\bibitem [{\citenamefont {Ludwig}\ and\ \citenamefont
  {Woodbury}(1961)}]{ludwig61}%
  \BibitemOpen
  \bibfield  {author} {\bibinfo {author} {\bibfnamefont {G.~W.}\ \bibnamefont
  {Ludwig}}\ and\ \bibinfo {author} {\bibfnamefont {H.~H.}\ \bibnamefont
  {Woodbury}},\ }\href {https://doi.org/10.1103/PhysRevLett.7.240} {\bibfield
  {journal} {\bibinfo  {journal} {Physical Review Letters}\ }\textbf {\bibinfo
  {volume} {7}},\ \bibinfo {pages} {240} (\bibinfo {year} {1961})}\BibitemShut
  {NoStop}%
\bibitem [{\citenamefont {Ham}(1961)}]{ham61}%
  \BibitemOpen
  \bibfield  {author} {\bibinfo {author} {\bibfnamefont {F.~S.}\ \bibnamefont
  {Ham}},\ }\href {https://doi.org/10.1103/PhysRevLett.7.242} {\bibfield
  {journal} {\bibinfo  {journal} {Physical Review Letters}\ }\textbf {\bibinfo
  {volume} {7}},\ \bibinfo {pages} {242} (\bibinfo {year} {1961})}\BibitemShut
  {NoStop}%
\bibitem [{\citenamefont {Bloembergen}(1961{\natexlab{b}})}]{bloembergen61a}%
  \BibitemOpen
  \bibfield  {author} {\bibinfo {author} {\bibfnamefont {N.}~\bibnamefont
  {Bloembergen}},\ }\href {https://doi.org/10.1103/PhysRevLett.7.90} {\bibfield
   {journal} {\bibinfo  {journal} {Physical Review Letters}\ }\textbf {\bibinfo
  {volume} {7}},\ \bibinfo {pages} {90} (\bibinfo {year}
  {1961}{\natexlab{b}})}\BibitemShut {NoStop}%
\bibitem [{\citenamefont {Armstrong}\ \emph {et~al.}(1961)\citenamefont
  {Armstrong}, \citenamefont {Bloembergen},\ and\ \citenamefont
  {Gill}}]{armstrong61}%
  \BibitemOpen
  \bibfield  {author} {\bibinfo {author} {\bibfnamefont {J.}~\bibnamefont
  {Armstrong}}, \bibinfo {author} {\bibfnamefont {N.}~\bibnamefont
  {Bloembergen}},\ and\ \bibinfo {author} {\bibfnamefont {D.}~\bibnamefont
  {Gill}},\ }\href {https://doi.org/10.1103/PhysRevLett.7.11} {\bibfield
  {journal} {\bibinfo  {journal} {Physical Review Letters}\ }\textbf {\bibinfo
  {volume} {7}},\ \bibinfo {pages} {11} (\bibinfo {year} {1961})}\BibitemShut
  {NoStop}%
\bibitem [{\citenamefont {Kushida}\ and\ \citenamefont
  {Saiki}(1961)}]{kushida61}%
  \BibitemOpen
  \bibfield  {author} {\bibinfo {author} {\bibfnamefont {T.}~\bibnamefont
  {Kushida}}\ and\ \bibinfo {author} {\bibfnamefont {K.}~\bibnamefont
  {Saiki}},\ }\href {https://doi.org/10.1103/PhysRevLett.7.9} {\bibfield
  {journal} {\bibinfo  {journal} {Physical Review Letters}\ }\textbf {\bibinfo
  {volume} {7}},\ \bibinfo {pages} {9} (\bibinfo {year} {1961})}\BibitemShut
  {NoStop}%
\bibitem [{\citenamefont {Kiel}(1966)}]{kiel66}%
  \BibitemOpen
  \bibfield  {author} {\bibinfo {author} {\bibfnamefont {A.}~\bibnamefont
  {Kiel}},\ }\href {https://doi.org/10.1103/PhysRev.148.247} {\bibfield
  {journal} {\bibinfo  {journal} {Physical Review}\ }\textbf {\bibinfo {volume}
  {148}},\ \bibinfo {pages} {247} (\bibinfo {year} {1966})}\BibitemShut
  {NoStop}%
\bibitem [{\citenamefont {{Hastings-Simon}}\ \emph {et~al.}(2006)\citenamefont
  {{Hastings-Simon}}, \citenamefont {Staudt}, \citenamefont {Afzelius},
  \citenamefont {Baldi}, \citenamefont {Jaccard}, \citenamefont {Tittel},\ and\
  \citenamefont {Gisin}}]{hastings-simon06}%
  \BibitemOpen
  \bibfield  {author} {\bibinfo {author} {\bibfnamefont {S.~R.}\ \bibnamefont
  {{Hastings-Simon}}}, \bibinfo {author} {\bibfnamefont {M.~U.}\ \bibnamefont
  {Staudt}}, \bibinfo {author} {\bibfnamefont {M.}~\bibnamefont {Afzelius}},
  \bibinfo {author} {\bibfnamefont {P.}~\bibnamefont {Baldi}}, \bibinfo
  {author} {\bibfnamefont {D.}~\bibnamefont {Jaccard}}, \bibinfo {author}
  {\bibfnamefont {W.}~\bibnamefont {Tittel}},\ and\ \bibinfo {author}
  {\bibfnamefont {N.}~\bibnamefont {Gisin}},\ }\href
  {https://doi.org/10.1103/PhysRevA.79.053851} {\bibfield  {journal} {\bibinfo
  {journal} {Optics Communications}\ }\textbf {\bibinfo {volume} {266}},\
  \bibinfo {pages} {716} (\bibinfo {year} {2006})}\BibitemShut {NoStop}%
\bibitem [{\citenamefont {Craiciu}(2020)}]{craiciu20}%
  \BibitemOpen
  \bibfield  {author} {\bibinfo {author} {\bibfnamefont {I.}~\bibnamefont
  {Craiciu}},\ }\emph {\bibinfo {title} {Quantum Storage of Light Using
  Nanophotonic Resonators Coupled to Erbium Ion Ensembles}},\ \href
  {https://doi.org/10.7907/yn6n-7x40} {Ph.D. thesis},\ \bibinfo  {school}
  {California Institute of Technology} (\bibinfo {year} {2020})\BibitemShut
  {NoStop}%
\bibitem [{\citenamefont {Mims}(1965)}]{mims65}%
  \BibitemOpen
  \bibfield  {author} {\bibinfo {author} {\bibfnamefont {W.~B.}\ \bibnamefont
  {Mims}},\ }\href {https://doi.org/10.1103/PhysRev.140.A531} {\bibfield
  {journal} {\bibinfo  {journal} {Physical Review}\ }\textbf {\bibinfo {volume}
  {140}},\ \bibinfo {pages} {A531} (\bibinfo {year} {1965})}\BibitemShut
  {NoStop}%
\bibitem [{\citenamefont {Kindem}\ \emph {et~al.}(2020)\citenamefont {Kindem},
  \citenamefont {Ruskuc}, \citenamefont {Bartholomew}, \citenamefont {Rochman},
  \citenamefont {Huan},\ and\ \citenamefont {Faraon}}]{kindem20}%
  \BibitemOpen
  \bibfield  {author} {\bibinfo {author} {\bibfnamefont {J.~M.}\ \bibnamefont
  {Kindem}}, \bibinfo {author} {\bibfnamefont {A.}~\bibnamefont {Ruskuc}},
  \bibinfo {author} {\bibfnamefont {J.~G.}\ \bibnamefont {Bartholomew}},
  \bibinfo {author} {\bibfnamefont {J.}~\bibnamefont {Rochman}}, \bibinfo
  {author} {\bibfnamefont {Y.~Q.}\ \bibnamefont {Huan}},\ and\ \bibinfo
  {author} {\bibfnamefont {A.}~\bibnamefont {Faraon}},\ }\href
  {https://doi.org/10.1038/s41586-020-2160-9} {\bibfield  {journal} {\bibinfo
  {journal} {Nature}\ }\textbf {\bibinfo {volume} {580}},\ \bibinfo {pages}
  {201} (\bibinfo {year} {2020})}\BibitemShut {NoStop}%
\bibitem [{\citenamefont {Liu}\ \emph {et~al.}(2020)\citenamefont {Liu},
  \citenamefont {Wang}, \citenamefont {Fang}, \citenamefont {Qin},
  \citenamefont {Wang}, \citenamefont {Jiang},\ and\ \citenamefont
  {Gao}}]{liu20}%
  \BibitemOpen
  \bibfield  {author} {\bibinfo {author} {\bibfnamefont {Z.}~\bibnamefont
  {Liu}}, \bibinfo {author} {\bibfnamefont {Y.-X.}\ \bibnamefont {Wang}},
  \bibinfo {author} {\bibfnamefont {Y.-H.}\ \bibnamefont {Fang}}, \bibinfo
  {author} {\bibfnamefont {S.-X.}\ \bibnamefont {Qin}}, \bibinfo {author}
  {\bibfnamefont {Z.-M.}\ \bibnamefont {Wang}}, \bibinfo {author}
  {\bibfnamefont {S.-D.}\ \bibnamefont {Jiang}},\ and\ \bibinfo {author}
  {\bibfnamefont {S.}~\bibnamefont {Gao}},\ }\href
  {https://doi.org/10.1093/nsr/nwaa148} {\bibfield  {journal} {\bibinfo
  {journal} {National Science Review}\ }\textbf {\bibinfo {volume} {7}},\
  \bibinfo {pages} {1557} (\bibinfo {year} {2020})}\BibitemShut {NoStop}%
\bibitem [{\citenamefont {Corna}\ \emph {et~al.}(2018)\citenamefont {Corna},
  \citenamefont {Bourdet}, \citenamefont {Maurand}, \citenamefont {Crippa},
  \citenamefont {{Kotekar-Patil}}, \citenamefont {Bohuslavskyi}, \citenamefont
  {Lavi{\'e}ville}, \citenamefont {Hutin}, \citenamefont {Barraud},
  \citenamefont {Jehl}, \citenamefont {Vinet}, \citenamefont {De~Franceschi},
  \citenamefont {Niquet},\ and\ \citenamefont {Sanquer}}]{corna18}%
  \BibitemOpen
  \bibfield  {author} {\bibinfo {author} {\bibfnamefont {A.}~\bibnamefont
  {Corna}}, \bibinfo {author} {\bibfnamefont {L.}~\bibnamefont {Bourdet}},
  \bibinfo {author} {\bibfnamefont {R.}~\bibnamefont {Maurand}}, \bibinfo
  {author} {\bibfnamefont {A.}~\bibnamefont {Crippa}}, \bibinfo {author}
  {\bibfnamefont {D.}~\bibnamefont {{Kotekar-Patil}}}, \bibinfo {author}
  {\bibfnamefont {H.}~\bibnamefont {Bohuslavskyi}}, \bibinfo {author}
  {\bibfnamefont {R.}~\bibnamefont {Lavi{\'e}ville}}, \bibinfo {author}
  {\bibfnamefont {L.}~\bibnamefont {Hutin}}, \bibinfo {author} {\bibfnamefont
  {S.}~\bibnamefont {Barraud}}, \bibinfo {author} {\bibfnamefont
  {X.}~\bibnamefont {Jehl}}, \bibinfo {author} {\bibfnamefont {M.}~\bibnamefont
  {Vinet}}, \bibinfo {author} {\bibfnamefont {S.}~\bibnamefont
  {De~Franceschi}}, \bibinfo {author} {\bibfnamefont {Y.-M.}\ \bibnamefont
  {Niquet}},\ and\ \bibinfo {author} {\bibfnamefont {M.}~\bibnamefont
  {Sanquer}},\ }\href {https://doi.org/10.1038/s41534-018-0059-1} {\bibfield
  {journal} {\bibinfo  {journal} {npj Quantum Information}\ }\textbf {\bibinfo
  {volume} {4}},\ \bibinfo {pages} {1} (\bibinfo {year} {2018})}\BibitemShut
  {NoStop}%
\bibitem [{\citenamefont {Krauth}\ \emph {et~al.}(2022)\citenamefont {Krauth},
  \citenamefont {Gorman}, \citenamefont {He}, \citenamefont {Jones},
  \citenamefont {Macha}, \citenamefont {Kocsis}, \citenamefont {Chua},
  \citenamefont {Voisin}, \citenamefont {Rogge}, \citenamefont {Rahman},
  \citenamefont {Chung},\ and\ \citenamefont {Simmons}}]{krauth22}%
  \BibitemOpen
  \bibfield  {author} {\bibinfo {author} {\bibfnamefont {F.}~\bibnamefont
  {Krauth}}, \bibinfo {author} {\bibfnamefont {S.}~\bibnamefont {Gorman}},
  \bibinfo {author} {\bibfnamefont {Y.}~\bibnamefont {He}}, \bibinfo {author}
  {\bibfnamefont {M.}~\bibnamefont {Jones}}, \bibinfo {author} {\bibfnamefont
  {P.}~\bibnamefont {Macha}}, \bibinfo {author} {\bibfnamefont
  {S.}~\bibnamefont {Kocsis}}, \bibinfo {author} {\bibfnamefont
  {C.}~\bibnamefont {Chua}}, \bibinfo {author} {\bibfnamefont {B.}~\bibnamefont
  {Voisin}}, \bibinfo {author} {\bibfnamefont {S.}~\bibnamefont {Rogge}},
  \bibinfo {author} {\bibfnamefont {R.}~\bibnamefont {Rahman}}, \bibinfo
  {author} {\bibfnamefont {Y.}~\bibnamefont {Chung}},\ and\ \bibinfo {author}
  {\bibfnamefont {M.}~\bibnamefont {Simmons}},\ }\href
  {https://doi.org/10.1103/PhysRevApplied.17.054006} {\bibfield  {journal}
  {\bibinfo  {journal} {Physical Review Applied}\ }\textbf {\bibinfo {volume}
  {17}},\ \bibinfo {pages} {054006} (\bibinfo {year} {2022})}\BibitemShut
  {NoStop}%
\bibitem [{\citenamefont {Osika}\ \emph {et~al.}(2022)\citenamefont {Osika},
  \citenamefont {Kocsis}, \citenamefont {Hsueh}, \citenamefont {Monir},
  \citenamefont {Chua}, \citenamefont {Lam}, \citenamefont {Voisin},
  \citenamefont {Simmons}, \citenamefont {Rogge},\ and\ \citenamefont
  {Rahman}}]{osika22}%
  \BibitemOpen
  \bibfield  {author} {\bibinfo {author} {\bibfnamefont {E.~N.}\ \bibnamefont
  {Osika}}, \bibinfo {author} {\bibfnamefont {S.}~\bibnamefont {Kocsis}},
  \bibinfo {author} {\bibfnamefont {Y.-L.}\ \bibnamefont {Hsueh}}, \bibinfo
  {author} {\bibfnamefont {S.}~\bibnamefont {Monir}}, \bibinfo {author}
  {\bibfnamefont {C.}~\bibnamefont {Chua}}, \bibinfo {author} {\bibfnamefont
  {H.}~\bibnamefont {Lam}}, \bibinfo {author} {\bibfnamefont {B.}~\bibnamefont
  {Voisin}}, \bibinfo {author} {\bibfnamefont {M.~Y.}\ \bibnamefont {Simmons}},
  \bibinfo {author} {\bibfnamefont {S.}~\bibnamefont {Rogge}},\ and\ \bibinfo
  {author} {\bibfnamefont {R.}~\bibnamefont {Rahman}},\ }\href
  {https://doi.org/10.1103/PhysRevApplied.17.054007} {\bibfield  {journal}
  {\bibinfo  {journal} {Physical Review Applied}\ }\textbf {\bibinfo {volume}
  {17}},\ \bibinfo {pages} {054007} (\bibinfo {year} {2022})}\BibitemShut
  {NoStop}%
\bibitem [{\citenamefont {Dajczgewand}\ \emph {et~al.}(2014)\citenamefont
  {Dajczgewand}, \citenamefont {Le~Gou{\"e}t}, \citenamefont
  {{Louchet-Chauvet}},\ and\ \citenamefont {Chaneli{\`e}re}}]{dajczgewand14}%
  \BibitemOpen
  \bibfield  {author} {\bibinfo {author} {\bibfnamefont {J.}~\bibnamefont
  {Dajczgewand}}, \bibinfo {author} {\bibfnamefont {J.-L.}\ \bibnamefont
  {Le~Gou{\"e}t}}, \bibinfo {author} {\bibfnamefont {A.}~\bibnamefont
  {{Louchet-Chauvet}}},\ and\ \bibinfo {author} {\bibfnamefont
  {T.}~\bibnamefont {Chaneli{\`e}re}},\ }\href
  {https://doi.org/10.1364/OL.39.002711} {\bibfield  {journal} {\bibinfo
  {journal} {Optics Letters}\ }\textbf {\bibinfo {volume} {39}},\ \bibinfo
  {pages} {2711} (\bibinfo {year} {2014})}\BibitemShut {NoStop}%
\bibitem [{\citenamefont {Sabooni}\ \emph {et~al.}(2016)\citenamefont
  {Sabooni}, \citenamefont {Nilsson}, \citenamefont {Kristensson},\ and\
  \citenamefont {Rippe}}]{sabooni16}%
  \BibitemOpen
  \bibfield  {author} {\bibinfo {author} {\bibfnamefont {M.}~\bibnamefont
  {Sabooni}}, \bibinfo {author} {\bibfnamefont {A.~N.}\ \bibnamefont
  {Nilsson}}, \bibinfo {author} {\bibfnamefont {G.}~\bibnamefont
  {Kristensson}},\ and\ \bibinfo {author} {\bibfnamefont {L.}~\bibnamefont
  {Rippe}},\ }\href {https://doi.org/10.1103/PhysRevA.93.013842} {\bibfield
  {journal} {\bibinfo  {journal} {Physical Review A}\ }\textbf {\bibinfo
  {volume} {93}},\ \bibinfo {pages} {013842} (\bibinfo {year}
  {2016})}\BibitemShut {NoStop}%
\end{thebibliography}%

\end{document}


\title{Supplemental material for: Effect of a hybrid transition moment on Stark-modulated photon echoes in Er\tplus:\yso}

\author{Rose L. \surname{Ahlefeldt}}
\email{rose.ahlefeldt@anu.edu.au}
\affiliation{Centre of Excellence for Quantum Computation and Communication Technology, Research School of Physics, Australian National University, Canberra, ACT 0200, Australia}
\author{Alexey \surname{Lyasota}}
\affiliation{Centre of Excellence for Quantum Computation and Communication Technology, School of Physics, University of New South Wales, Sydney, NSW 2052, Australia}
\author{Jodie \surname{Smith}}
\affiliation{Centre of Excellence for Quantum Computation and Communication Technology, Research School of Physics, Australian National University, Canberra, ACT 0200, Australia}
\author{Jinliang \surname{Ren}}
\affiliation{Centre of Excellence for Quantum Computation and Communication Technology, Research School of Physics, Australian National University, Canberra, ACT 0200, Australia}
\author{Matthew J. \surname{Sellars}}
\affiliation{Centre of Excellence for Quantum Computation and Communication Technology, Research School of Physics, Australian National University, Canberra, ACT 0200, Australia}
\maketitle

\section{ Optical Bloch equation model: reduced visibility}
 Here, we expand on the conditions in which a reduction of visibility of the Stark-modulated echo signal indicates a mixed transition dipole moment. This statement is true as long as the atom ensemble being driven by the optical pulses has an inhomogeneous linewidth much greater than the optical pulse bandwidth. This is the normal situation for Stark modulated echoes, since the inhomogeneous broadening of optical transitions is typically $>\SI{100}{\mega\hertz}$ and the pulse bandwidth is typically of the order of \SI{1}{\mega\hertz}.  However, it is possible to get a reduction in visibility even for an electric-dipole-only transition if two conditions are met: the linewidth of the atoms is less than the pulse bandwidth (for instance, if a narrow ensemble is selected with spectral holeburning), and the rephasing pulse in the echo sequence (the `$\pi$' pulse) is significantly more or less than a $\pi$ pulse. 

\begin{figure}
\begin{center}
    \includegraphics[width=\columnwidth]{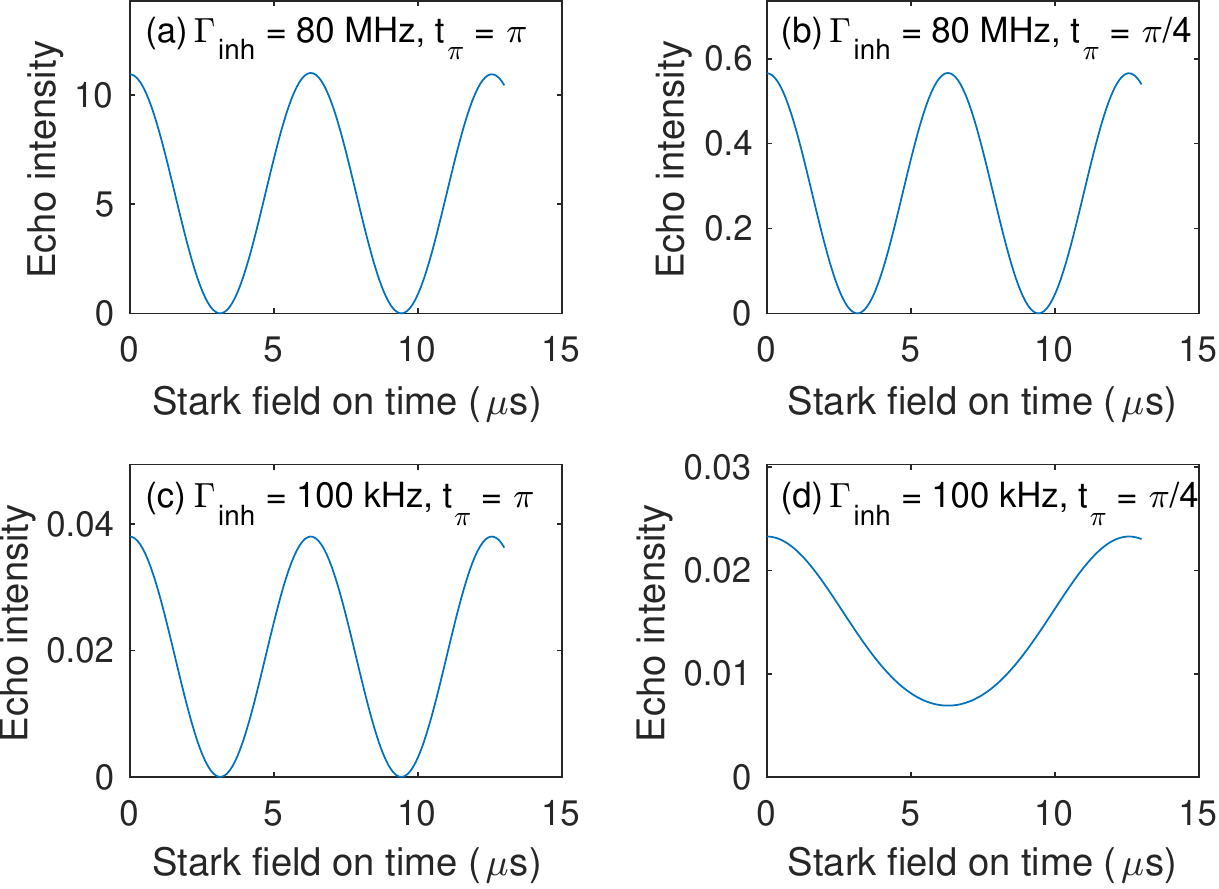}
    \caption{ \label{underdrive} Optical Bloch equation model of Stark modulated photon echo signal, for an electric-dipole-only transition, different ensemble inhomogeneous linewidths $\Gamma_{inh} = \SI{80}{\mega\hertz}$ natural inhomogeneous broadening), (a), (b)  and \SI{100}{\kilo\hertz} (spectrally holeburnt feature). We consider two lengths of echo pulses: $t_\pi=\pi$ and $t_\pi=\pi/4$.}
\end{center}
\end{figure}
 To demonstrate this effect, we use the optical Bloch model described in the main text to determine the Stark modulated echo signal for an electric-dipole only transition for four different regimes in Fig. \ref{underdrive}: a broad (80~MHz) ensemble for a perfect and for a weak $\pi$-pulse ($t_\pi = \pi$ and $t_\pi=\pi/4$ respectively), and a narrow (100~kHz) ensemble for the same perfect and weak pulses. In all cases the Rabi frequency is 1~MHz and the pulses are applied in $x$. 

 For both ensemble widths, the perfect $\pi$ pulse gives a 100\% visibility modulation, as does the weak pulse for the broad ensemble, but for the narrow ensemble it produces a modulation with half the frequency of the other figures, and visibility well below 100\%. This is because the weak pulse takes the Stark frequency shift encoded into the phase of the Bloch vector, and maps it into a state confined to the $+y$ hemisphere of the Bloch sphere. If there is no Stark on time for which the state crosses the $x$-axis, the echo signal will never reach zero. This effect is mitigated in the broad ensemble, because the wide range of generalised Rabi frequencies of the ions means the Bloch vectors of the ensemble are always distributed over both hemispheres, regardless of the pulse length. 
 
 It is straightforward to distinguish a reduced visibility due to underdriving (or overdriving) a narrow distribution from reduced visibility from a hybrid moment: a narrow distribution must have been prepared, and the signal seen will be highly dependent on the pulse length. If a reduced visibility is seen outside of this special case, or if it is accompanied by phase and polarisation shifts of the signal, it indicates the presence of a mixed transition dipole moment.  
 
\section{Two-pulse echo decays}
\begin{figure}
\begin{center}
    \includegraphics[width=\columnwidth]{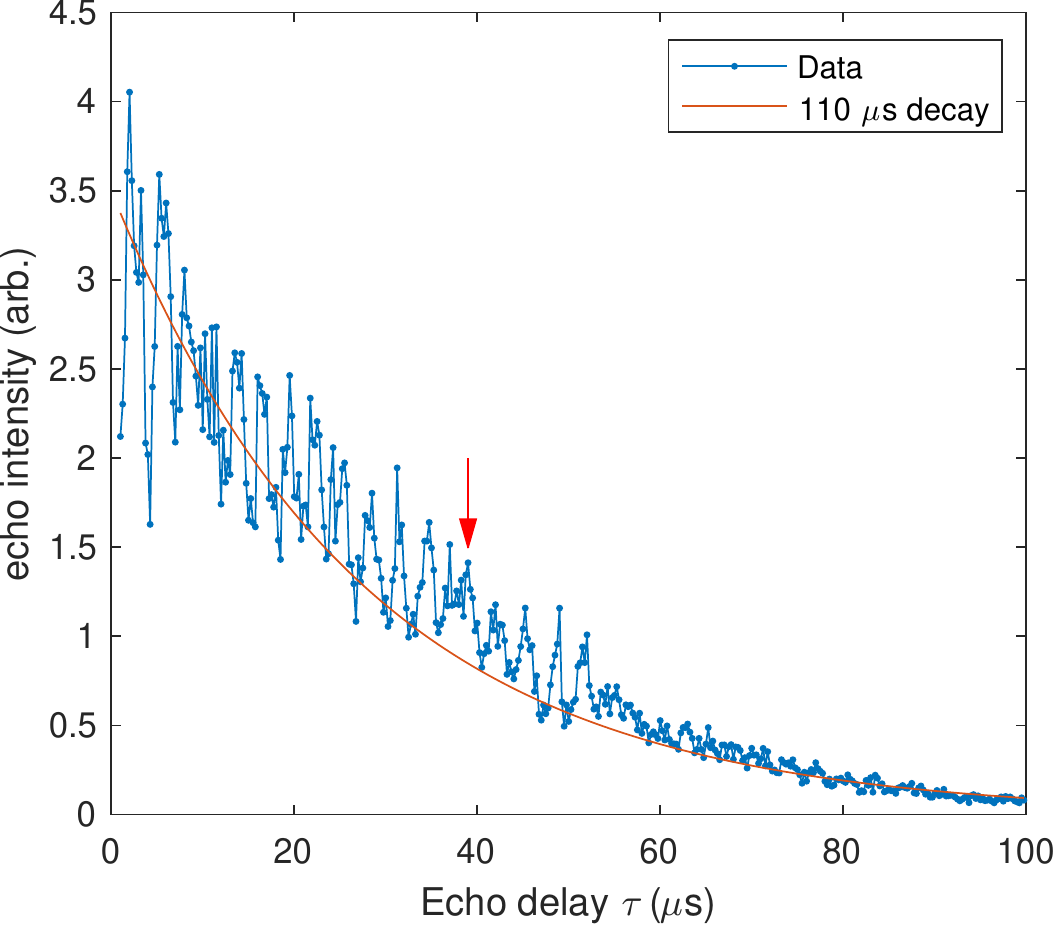}
    \caption{ \label{echomod} Example of superhyperfine modulations in a two-pulse echo decay, here for site 2 and D$_1$ polarisation, C$_2$ propagation, taken with zero electric field. The overall decay time is $T_2\approx 110$~\unit{\micro\second} as indicated by the orange line. The red arrow indicates the delay time chosen for Stark-modulated echo measurements in this direction.}
\end{center}
\end{figure}
As discussed in the main text, at the 300~mT magnetic field here, the \ert spin is still coupled to the nuclear spin bath, producing superhyperfine modulations in a two-pulse echo decay sequence, as seen in Fig. \ref{echomod} below. To maximize the signal for Stark-modulated echoes, we mapped these modulations in zero electric field and then chose a delay for the Stark-modulated measurements where the interference due to these modulations was minimal. For electric fields in the D$_1$ and D$_2$ directions of site 2, where the Stark shifts were weak, delays of the order of \SI{40}{\micro\second} were used, while for all other measurements, delays of the order of \SI{20}{\micro\second} were used.